\documentclass[twocolumn]{IEEEtran}
\IEEEoverridecommandlockouts
\usepackage{cite}
\usepackage{textcomp}
\usepackage{xcolor}
\usepackage[mathscr]{eucal}
\usepackage[cmex10]{amsmath}
\usepackage{epsfig,epsf,psfrag}
\usepackage{amssymb,amsmath,amsthm,amsfonts,latexsym}
\usepackage{amsmath,graphicx,bm,xcolor,url,overpic}
\usepackage{subfigure}
\usepackage{fixltx2e}
\usepackage{array}
\usepackage{verbatim}
\usepackage{bm}
\usepackage{bbm}
\usepackage{algorithmic}
\usepackage{algorithm}
\usepackage{verbatim}
\usepackage{textcomp}
\usepackage{mathrsfs}
\usepackage{epstopdf}

\newcommand{\openone}{\leavevmode\hbox{\small1\normalsize\kern-.33em1}}

\catcode`~=11 \def\UrlSpecials{\do\~{\kern -.15em\lower .7ex\hbox{~}\kern .04em}} \catcode`~=13 

\allowdisplaybreaks[1]

\newcommand{\nn}{\nonumber}

\newcommand{\calB}{\mathcal{B}}

\newcommand{\calE}{\mathcal{E}}
\newcommand{\calF}{\mathcal{F}}
\newcommand{\calG}{\mathcal{G}}

\newcommand{\calN}{\mathcal{N}}

\newcommand{\calP}{\mathcal{P}}

\newcommand{\calT}{\mathcal{T}}

\newcommand{\calX}{\mathcal{X}}

\newcommand{\be}{\mathbf{e}}

\newcommand{\bg}{\mathbf{g}}

\newcommand{\bH}{\mathbf{H}}

\newcommand{\bJ}{\mathbf{J}}

\newcommand{\bbE}{\mathbb{E}}

\newcommand{\bbN}{\mathbb{N}}

\newcommand{\bbR}{\mathbb{R}}

\DeclareMathAlphabet{\mathbsf}{OT1}{cmss}{bx}{n}
\DeclareMathAlphabet{\mathssf}{OT1}{cmss}{m}{sl}

\newcommand{\rvP}{\mathsf{P}}

\DeclareSymbolFont{bsfletters}{OT1}{cmss}{bx}{n}  
\DeclareSymbolFont{ssfletters}{OT1}{cmss}{m}{n}
\DeclareMathSymbol{\bsfGamma}{0}{bsfletters}{'000}
\DeclareMathSymbol{\ssfGamma}{0}{ssfletters}{'000}
\DeclareMathSymbol{\bsfDelta}{0}{bsfletters}{'001}
\DeclareMathSymbol{\ssfDelta}{0}{ssfletters}{'001}
\DeclareMathSymbol{\bsfTheta}{0}{bsfletters}{'002}
\DeclareMathSymbol{\ssfTheta}{0}{ssfletters}{'002}
\DeclareMathSymbol{\bsfLambda}{0}{bsfletters}{'003}
\DeclareMathSymbol{\ssfLambda}{0}{ssfletters}{'003}
\DeclareMathSymbol{\bsfXi}{0}{bsfletters}{'004}
\DeclareMathSymbol{\ssfXi}{0}{ssfletters}{'004}
\DeclareMathSymbol{\bsfPi}{0}{bsfletters}{'005}
\DeclareMathSymbol{\ssfPi}{0}{ssfletters}{'005}
\DeclareMathSymbol{\bsfSigma}{0}{bsfletters}{'006}
\DeclareMathSymbol{\ssfSigma}{0}{ssfletters}{'006}
\DeclareMathSymbol{\bsfUpsilon}{0}{bsfletters}{'007}
\DeclareMathSymbol{\ssfUpsilon}{0}{ssfletters}{'007}
\DeclareMathSymbol{\bsfPhi}{0}{bsfletters}{'010}
\DeclareMathSymbol{\ssfPhi}{0}{ssfletters}{'010}
\DeclareMathSymbol{\bsfPsi}{0}{bsfletters}{'011}
\DeclareMathSymbol{\ssfPsi}{0}{ssfletters}{'011}
\DeclareMathSymbol{\bsfOmega}{0}{bsfletters}{'012}
\DeclareMathSymbol{\ssfOmega}{0}{ssfletters}{'012}

\newcommand{\bgamma}{\bm{\gamma}}

\newcommand{\veps}{\varepsilon}

\DeclareMathOperator*{\argmin}{arg\,min}

\usepackage{thmtools, thm-restate}
\newtheorem{theorem}{Theorem} 
\newtheorem{lemma}[theorem]{Lemma}

\newtheorem{proposition}[theorem]{Proposition}

\newtheorem{example}{Example}

\usepackage{cite}
\usepackage{amsmath,amssymb,amsfonts,bbm}
\usepackage{algorithmic}
\usepackage{graphicx}
\usepackage{textcomp}
\usepackage{xcolor,overpic}
\usepackage{hyperref}
\hypersetup{
    colorlinks,
    linkcolor = {red!60!black},
    citecolor = {blue!60!black},
    urlcolor = {blue!50!black}
}
\usepackage[justification=centering]{caption}
\def\BibTeX{{\rm B\kern-.05em{\sc i\kern-.025em b}\kern-.08em
		T\kern-.1667em\lower.7ex\hbox{E}\kern-.125emX}}

\newtheoremstyle{noparens}%
{}{}%
{\itshape}{}%
{\bfseries}{.}%
{ }%
{\thmname{#1}\thmnumber{ #2}\mdseries\thmnote{ #3}}
	
\theoremstyle{noparens}

\begin{document}
	
	\title{Asymptotics of Sequential Composite Hypothesis Testing under Probabilistic Constraints\\
	}
	
	\author{\IEEEauthorblockN{Jiachun Pan, Yonglong Li, Vincent Y.~F.~Tan, {\em Senior Member, IEEE}}   \thanks{This work is partially funded by a Singapore National Research Foundation Fellowship (R-263-000-D02-281). The paper was presented in part at the 2021 International Symposium on Information Theory (ISIT) \cite{pan2021}. 
	
	Jiachun Pan is with the Department of Electrical and Computer Engineering, National University of Singapore, Singapore, Email: \url{pan.jiachun@u.nus.edu}. Yonglong Li is with the Department of Electrical and Computer Engineering, National University of Singapore, Singapore, Email: \url{elelong@nus.edu.sg}.
	Vincent Y.~F.~Tan is with the Department of Electrical and Computer Engineering and the Department of Mathematics, National University of Singapore, Singapore, Email: \url{vtan@nus.edu.sg}.
	
	Copyright (c) 2017 IEEE. Personal use of this material is permitted.  However, permission to use this material for any other purposes must be obtained from the IEEE by sending a request to \url{pubs-permissions@ieee.org}.
	}   
		
		
		
	}

	\maketitle
	
	\begin{abstract}
		We consider the sequential composite binary hypothesis testing problem in which one of the hypotheses is governed by a single distribution while the other is governed by a family of distributions whose parameters belong to a known set $\Gamma$. We would like to design a test to decide which hypothesis is in effect. Under the constraints that the probabilities that the length of the test, a stopping time, exceeds $n$ are bounded by a certain threshold $\epsilon$, we obtain certain fundamental limits on the asymptotic behavior of the sequential test as $n$ tends to infinity. Assuming that $\Gamma$ is a convex and compact set, we obtain the set of all first-order error exponents for the problem. We also prove a strong converse. Additionally, we obtain the set of second-order error exponents under the assumption that the alphabet of the observations $\calX$ is finite. In the proof of second-order asymptotics, a main technical contribution is the derivation of a central limit-type result for a maximum of an uncountable set of log-likelihood ratios under suitable conditions. This result may be of independent interest.  We also show that some important statistical models  satisfy the conditions.
	\end{abstract}
	
	
	\section{Introduction}
	Hypothesis testing is a fundamental problem in information theory and statistics~\cite{blahut1974hypothesis}. Here we consider a sequential composite hypothesis testing problem in which i.i.d. observations are drawn from either a simple null hypothesis or a composite from the alternative hypothesis. We consider the first-order and second-order tradeoff between the two types of error probabilities under a probabilistic constraint on the stopping times. There is a vast literature on this subject~\cite[Part I]{tartakovsky2014sequential}, however the optimal trade-off in the probabilistic stopping time constraint has not been determined. The probabilistic constraints means that we constrain the  probabilities (under both hypotheses) that the length of the stopping time exceeds $n$ to be no larger than some prescribed threshold $\epsilon \in (0,1)$. We let $n$ tend to infinity to exploit various asymptotic and concentration results.

	\subsection{Related works}
	\label{sec:related}
 In the classical problem of sequential hypothesis testing in the statistical literature, one seeks to minimize the expected sample size $\mathbb{E}_i[\tau(\delta)],i\in\{0,1\}$ subject to bounds on the type-I and type-II error probabilities $P_0(\delta_{\tau}=1)\leq \alpha$ and $P_1(\delta_{\tau}=0)\leq \beta$, i.e., the sequential testing problem is to solve, for each $i\in\{0,1\}$,
	\begin{align}
		\label{eqn:dual}
		\min_{(\tau,\delta)} \mathbb{E}_i[\tau(\delta)] \quad \mbox{s.t.~}\quad P_0(\delta_{\tau}=1)\leq \alpha,P_1(\delta_{\tau}=0)\leq \beta.
	\end{align} 
	There is a vast literature on solving the above problem (see~\cite[Part I]{tartakovsky2014sequential} for example). The dual problem corresponding to that of~\eqref{eqn:dual} is the minimization of the error probabilities subject to expectation constraints on the sample size. More specifically, the dual problem corresponding to~\eqref{eqn:dual} entails solving for each $i\in\{0,1\}$,
	\begin{align}\label{eqn:dual1}
		\min_{(\tau,\delta)} P_i(\delta_{\tau}=1-i) \quad \mbox{s.t.~} \quad\mathbb{E}_i[\tau]\leq n,i\in\{0,1\}.
	\end{align}
	The optimal tests $(\tau^*,\delta^*)$ of~\eqref{eqn:dual} and~\eqref{eqn:dual1} are given by appropriate sequential probability ratio tests. However, in this paper, we consider the problem of minimizing the error probabilities subject to {\em probabilistic  constraints on the sample size}. In more detail, the problem we are concerned with is the following:
	\begin{align}\label{eqn:prob}
		\min_{(\tau,\delta)}  P_i(\delta_{\tau}=1-i) \quad \mbox{s.t.~}\quad P_i(\tau>n)<\epsilon,i\in\{0,1\}.
\end{align}
As the nature of the constraints are different (expectation versus probabilistic), the proof techniques are also different. For problem~\eqref{eqn:dual1}, Wald's identity and data-processing inequality are used to derive the achievability and the converse. For our problem~\eqref{eqn:prob}, concentration inequalities such as the central limit theorem  are used to derive the achievability and the converse.

For the first-order asymptotics (exponents of the two types of error probabilities), there is a vast literature on binary hypothesis testing. In the fixed-length hypothesis testing where the length of the vector of observations is fixed, the Neyman--Pearson lemma~\cite{neyman1933ix} states that the likelihood ratio test is optimal and Chernoff--Stein lemma~\cite[Theorem~13.1]{polyanskiy2014lecture} shows that if we constrain the type-I error to be less than any $\epsilon\in(0,1)$, the best (maximum) type-II error exponent is the relative entropy $D(p_0\|p_1)$,  where $p_0$ and $p_1$ are respectively the distributions under the null and alternative hypotheses respectively. If we require the type-I error exponent  to be at least $r>0$, i.e., the type-I error probability is upper bounded by $\exp(-nr)$,  the maximum  type-II error exponent   is $\min\{ D(q\|p_0) : D(q\|p_1)\ge r\}$~\cite{blahut1974hypothesis}. In this regard, we see that there is a trade-off between two error exponents, i.e., they cannot be simultaneously large.  However, in the {\em sequential} case where the length of the test sample is a stopping time and its expectation is bounded by $n$, the trade-off can be eradicated. Wald and Wolfowitz~\cite{wald1948} showed that when the expectations of sample length under $H_0$ and $H_1$ are bounded by a common integer $n$ (these are known as the expectation constraints) and $n$ tends to infinity, the set of achievable error exponents is $\{(E_0,E_1):E_0\le  D(p_1\|p_0), E_1\leq D(p_0\|p_1)\}$. In addition, the corner point $(D(p_1\|p_0),D(p_0\|p_1))$ is attained by  a sequence of sequential probability ratio tests (SPRTs). Lalitha and Javidi~\cite{lalitha2016reliability} considered an interesting setting that interpolates between the fixed-length hypothesis testing and sequential hypothesis testing. They considered the {\em almost-fixed-length hypothesis testing} problem, in which the stopping time is allowed to be larger than a prescribed integer $n$ with exponentially small probability $\exp(-n\gamma)$ for different $\gamma>0$. The probabilistic constraints we employ in this paper are analogous to those in~\cite{lalitha2016reliability}, but instead of allowing the event that the stopping time to be larger than $n$ to have exponentially small probability, we only require this event to have probability at most $\epsilon\in(0,1)$, a fixed constant. This allows us to ask questions ranging from strong converses to second-order asymptotics. In \cite{haghifam}, Haghifam, Tan, and Khisti considered {\em sequential classification} which is similar to sequential hypothesis testing apart from the fact that true distributions are only partially known in the form of training samples. 
	
	For the composite hypothesis testing, Zeitouni, Ziv, and Merhav~\cite{zeitouni} investigated the generalized likelihood ratio test (GLRT) and proposed conditions for asymptotic optimality of the GLRT in the Neyman-Pearson sense. For the sequential case, Lai~\cite{lai2002asymptotic} analyzed different sequential testing problems and obtained a unified asymptotic theory that results in certain generalized sequential likelihood ratio tests to be asymptotically optimal solutions to these problem.   Li, Nitinawarat and Veeravalli~\cite{Li14} considered a universal outlier hypothesis testing problem in the fixed-length setting; {\em universality} here refers to the fact that the distributions are unknown and have to be estimated on the fly.   They then extended their work to the sequential  setting~\cite{Li17} but under expectation constraints on the stopping time.  The work that is closest to ours is that by  Li, Liu, and Ying~\cite{Xiaoou} whose results can be  modified to solve the composite version of the dual problem~\eqref{eqn:dual1}. They showed that the generalized sequential probability ratio test is asymptotically optimal by making use of optimality results of sequential probability ratio tests (SPRTs).

	Concerning the second-order asymptotic regime, in fixed-length binary hypothesis testing in which the type-I error is bounded by a fixed constant $\epsilon\in (0,1)$, Strassen~\cite{strassen1962asymptotische} showed that the second-order term can be quantified via the relative entropy variance~\cite{tan2014asymptotic} and the inverse of the Gaussian cumulative distribution function. For the sequential case, Li and Tan~\cite{yonglong} recently established the second-order asymptotics of sequential binary hypothesis testing under probabilistic and expectation constraints on the stopping time, showing that the former (resp.\ latter) set of constraints results in a $\Theta(1/\sqrt{n})$ (resp.\ $\Theta(1/n)$) backoff from the relative entropies. These are estimates of the costs of operating in the finite-length setting.  In this paper, we seek to extend these results to sequential {\em composite} hypothesis testing.
	
	\subsection{Main contributions}
	\label{sec:maincontribution}
	Our main contributions consist in obtaining the first-order and second-order asymptotics for sequential composite hypothesis testing under the probabilistic constraints, i.e., we constrain the probabilities that the lengths of observations exceed $n$ is no larger than some prescribed $\epsilon\in (0,1)$. 
\begin{itemize}
\item 	First,  while the results of Li, Liu, and Ying~\cite{Xiaoou} can be modified to solve the composite version of the dual problem in~\eqref{eqn:dual1}, which yields first-order asymptotic results under expectation constraints, we obtain the first-order asymptotic results under the probabilistic constraints. We show that the corner points of the optimal error exponent regions are identical under both types of constraints. 

\item  Second,  Li, Liu, and Ying~\cite{Xiaoou} only proved that the generalized sequential probability ratio test is asymptotically optimal by making use of the optimality results of sequential probability ratio test (SPRT). Here we prove a {\em strong converse} result, namely that the exponents stay unchanged even if the probability that the stopping time exceeds $n$ is smaller than $\epsilon$ for all $\epsilon\in(0,1)$. We do so using information-theoretic ideas and, in particular,  the ubiquitous change-of-measure technique (Lemma~\ref{lemma:converse}). 
\item Third, and most importantly, we obtain the second-order asymptotics of the error exponents when we assume that the observations take values on a finite alphabet. A main technical contribution here is that we obtain a new central limit-type result for a maximum of an uncountable set of log-likelihood ratios under suitable conditions (Proposition~\ref{lem:gaussian}). We contrast our central limit-type result to classical statistical results such as Wilks' theorem~\cite[Chapter 16]{vaart_1998}.
	
\end{itemize}	
	
\subsection{Paper Outline}
The rest of the paper is structured as follows.  In Section~\ref{sec:formulation}, we formulate the composite sequential hypothesis testing problem precisely and state the probabilistic constraints on the stopping time. In Section~\ref{first-order}, we list some mild assumptions on the distributions and uncertainty set in order to state and prove our first-order asymptotic results.  In Section~\ref{second-order}, we consider the second-order asymptotics of the same problem by augmenting to the assumptions stated in Section~\ref{first-order}. We state a central limit-type theorem for the maximum of a set of log-likelihood ratios and our main result concerning the second-order asymptotics. We  relegate the more technical calculations (such as proofs of lemmata) to the appendix. 
	
	\section{Problem Formulation} \label{sec:formulation}
	
	Let $\{X_i\}_{i=1}^{\infty}$ be an observed i.i.d.\ sequence, where each $X_i$ follows a density $p$ with respect to a base measure $\mu$ on the alphabet~$\calX$. We consider the problem of composite hypothesis testing:
	\begin{align*}
	H_0: p=p_0 \quad
 \mbox{and}\quad H_1: p\in\{p_{\gamma}:\gamma\in\Gamma\},
	\end{align*}
	where $p_0$ and $p_{\gamma}$ are density functions with respect to $\mu$ and $p_0\notin\{p_{\gamma}:\gamma\in\Gamma\}$. We assume that $p_0$ and $p_{\gamma}$ are mutually absolutely continuous for all $\gamma\in\Gamma$.  Denote $P_0$ and $P_{\gamma}$ as the probability measures associated to $p_0$ and $p_{\gamma}$, respectively. Let $\calF(X^n)$ be the $\sigma$-algebra generated by $X^n=(X_1,X_2,\ldots, X_n)$. Let $\tau$ be a stopping time adapted to the filtration $\{\calF(X^n)\}_{n=1}^{\infty}$ and let $\calF_{\tau}$ be the $\sigma$-algebra associated with $\tau$. Let $\delta$ be a $\{0,1\}$-valued $\calF_{\tau}$-measurable function. The pair $(\delta,\tau)$ constitutes a {\em sequential hypothesis test}, where $\delta$ is called the \emph{decision function} and $\tau$ is the {\em stopping time}. When $\delta=0$ (resp.~$\delta=1$), the decision is made in favor of $H_0$ (resp.~$H_1$). The \emph{type-I} and \emph{maximal type-II error probabilities} are defined as
	\begin{align*}
	\rvP_{1|0}(\delta,\tau):=P_{0}(\delta=1)\quad
 \mbox{and}\quad \rvP_{0|1}(\delta,\tau):=\sup_{\gamma\in\Gamma}P_{\gamma}(\delta=0).
	\end{align*}
	In other words, $\rvP_{1|0}(\delta,\tau)$ is the error probability that the true density is $p_0$ but $\delta=1$ and $\rvP_{0|1}(\delta,\tau)$ is the maximal error probability over all parameters $\gamma\in\Gamma$ that the true density is $p_\gamma$ but the decision made $\delta=0$ based on the observations up to time $\tau$. 
	
	In this paper, we seek the first-order and second-order asymptotics of exponents of the error probabilities under \emph{probabilistic constraints} on stopping time $\tau$. The probabilistic constraints dictate that, for every error tolerance $0<\epsilon<1$, there exists an integer~$n_0(\epsilon)$ such that for all $n>n_0(\epsilon)$, the stopping time $\tau$ satisfies   
	\begin{align}
	\label{eqn:probcons}
	P_{0}(\tau>n)<\epsilon\quad\mbox{and} \quad \sup_{\gamma\in\Gamma}P_{\gamma}(\tau>n)<\epsilon,
	\end{align}
 and 
\begin{align}
\label{eqn:probcons2}
P_0(\tau<\infty)=1 \quad \mbox{and}\quad \sup_{\gamma\in\Gamma}P_{\gamma}(\tau<\infty)=1.
\end{align}
	In the following, all logarithms are natural logarithms, i.e., with respect to base $e$.

	\section{First-order Asymptotics}
	\label{first-order}
	We say that an exponent pair $(E_0,E_1)$ is {\em $\epsilon$-achievable under the probabilistic constraints} if there exists a sequence of sequential hypothesis tests $\{(\delta_n,\tau_n)\}_{n=1}^{\infty}$ that satisfies the probabilistic constraints on the stopping time in~\eqref{eqn:probcons} and~\eqref{eqn:probcons2} and
	\begin{align*}
	E_0&\leq\liminf_{n\to\infty}\frac{1}{n}\log\frac{1}{\rvP_{1|0}(\delta_n,\tau_n)},\\
	 E_1&\leq\liminf_{n\to\infty}\frac{1}{n}\log\frac{1}{\rvP_{0|1}(\delta_n,\tau_n)}.
	\end{align*}
	The set of all $\epsilon$-achievable $(E_0,E_1)$ is denoted as $\calE_{\epsilon}(p_{0},\Gamma)$. For simple (non-composite) binary sequential hypothesis testing under the expectation constraints (i.e., $\max_{i=0,1}\mathbb{E}_{P_i}[\tau]\leq n)$, the set of all achievable error exponent pairs, as shown by Wald and Wolfowitz~\cite{wald1948} (also see \cite{yonglong,lalitha2016reliability}), is 
	\begin{align}
	\tilde{\calE}_{\epsilon}(p_0,p_1)=\{(E_0,E_1):E_0\le  D(p_1\|p_0), E_1\leq D(p_0\|p_1)\}.\label{eqn:region_expectation}
	\end{align}
	The corner point $(D(p_1\|p_0),D(p_0\|p_1))$ can be achieved by a sequence of sequential probability ratio tests~\cite{wald1948}. 
	
	We define the {\em log-likelihood ratio} and {\em maximal log-likelihood ratio} respectively as 
	\begin{align*}
	S_n(\gamma):= \sum_{i=1}^n \log\frac{p_{\gamma}(X_i)}{p_{0}(X_i)}\quad \mbox{and}\quad S_n:=\sup_{\gamma\in\Gamma}S_n(\gamma).
	\end{align*}
	For two positive numbers $A$ and $B$, we define the stopping time $\tau$ as
	\begin{align*}
	\tau:=\inf\{n:S_n>A \text{ or } S_n<-B\},
	\end{align*} 
	and the decision rule as
	\begin{align*}
	\delta:=
	\begin{cases}
	 0,& \text{if } S_{\tau}<-B,\\
	 1, &\text{if } S_{\tau}>A.
	\end{cases}
	\end{align*}
	We term the above test given by $(\delta,\tau)$ as a {\em generalized sequential probability ratio test} (GSPRT) with thresholds $A$ and $B$. The stopping time $\tau$ is almost surely finite for any distribution within the family~\cite{Xiaoou}, so~\eqref{eqn:probcons2} holds for GSPRT.
	For the above GSPRT, we define {\em type-I error probability} and {\em maximal type-II error probability} respectively as
	\begin{align*}
	\rvP_{1|0}(\tau,\delta)&:=P_{0}(S_{\tau}>A),~\mbox{and}\\ \rvP_{0|1}(\tau,\delta)&:=\sup_{\gamma\in\Gamma}P_{\gamma}(S_{\tau}<-B).
	\end{align*}
	We   introduce some assumptions on the distributions and~$\Gamma$.
	\begin{itemize}
		\item[(A1)] The parameter set $\Gamma\subset\bbR^d$ is compact.

		\item[(A2)] Assume that  $\gamma\mapsto D(p_{\gamma}\|p_0)$ and $\gamma\mapsto D(p_0\|p_{\gamma})$ are    twice continuously differentiable on $\Gamma$.  For each $\gamma\in\Gamma$, the solutions to the minimizations $\min_{\gamma\in\Gamma}D(p_0\|p_{\gamma})$ and $\min_{\gamma\in\Gamma}D(p_{\gamma}\|p_0)$  are unique. Their existences are guaranteed by the compactness  of $\Gamma $ and the continuity of $\gamma\mapsto D(p_{\gamma}\|p_0)$ and $\gamma \mapsto D(p_0\|p_{\gamma})$ on $\Gamma$.    In addition,
		$\min_{\gamma\in\Gamma}D(p_0\|p_{\gamma})>\epsilon_0$ and $\min_{\gamma\in\Gamma}D(p_{\gamma}\|p_0)>\epsilon_0$ for some $\epsilon_0>0$. 
		
		\item[(A3)]		Let $\xi(\gamma)=\log p_{{\gamma}}(X)-\log p_{0}(X)$ be the log-likelihood ratio. We assume that  $\mathbb{E}[\max_{\gamma}|\xi(\gamma)|^2]<\infty$.
		Besides, there exist~$\alpha>1$ and~$x_0\in\mathbb{R}$ such that for all $\gamma\in\Gamma$, and $x>x_0$
		\begin{align}
		\label{eqn:con}
		P_{0}\Bigg(\max_{\gamma\in\Gamma}|\nabla_{\gamma}\xi(\gamma)|>x\Bigg)\leq e^{-|\log x|^{\alpha}},
		\end{align}
		where    $|\nabla_{\gamma}\xi(\gamma)|$ is the $\ell_1$ norm of the gradient vector $\nabla_{\gamma}\xi(\gamma)\in\bbR^d$
	\end{itemize}
	We present some examples that satisfy Conditions (A1)--(A3).   We first show that Condition (A1)--(A3) hold for the canonical exponential family under suitable assumptions and then provide an explicit example. 
	\begin{example}[Canonical exponential families] {\em  
	The general form of probability density for the canonical exponential family of probability distributions is ~\cite{wainwright2008graphical}:
			\begin{align*}
				p_{\bgamma}(x)=h(x)\exp(\bgamma^\top T(x)-A(\bgamma)),
			\end{align*}
			where $h(x)$ is called the base measure, $\bgamma$ is the parameter vector, $T(x)$ is referred to as the sufficient statistic and $A(\bgamma)$ is the cumulant generating function.	We define the set of valid parameters as $\Theta=\{\bgamma\in\bbR^d:A(\bgamma)<\infty\}$.
			
			Now we consider the test
			\begin{align*}
				&H_0: p_{\bgamma_0}(x)=h(x) \exp(\bgamma_0^\top T(x)-A(\bgamma_0)), \quad\bgamma_0\in\Theta; \\
				& H_1: p_{\bgamma}(x)=h(x) \exp(\bgamma^\top T(x)-A(\bgamma)), \quad\bgamma\in\Gamma, \bgamma_0\notin\Gamma.
			\end{align*}
			We also assume that the exponential families under consideration satisfy the following assumptions: 
			\begin{enumerate}
				\item [(i)] $\Gamma\subset\Theta$ is a convex and compact set;
				\item [(ii)] $A(\bgamma)$ is thrice continuously differentiable with respect to $\bgamma$;
				\item [(iii)] $\nabla_{\bgamma}^2 A(\bgamma)$ and $\nabla((\bgamma-\bgamma_0)^\top \nabla_{\bgamma}^2 A(\bgamma))$ are positive definite for $\bgamma\in\Gamma$.
			\end{enumerate}

			For this example, Condition (A1) holds because of Assumption (i). For Condition (A2), we have 
			\begin{align*}
				D(p_{\bgamma}\|p_{\bgamma_0})=(\bgamma-\bgamma_0)^\top\mathbb{E}_{\bgamma}[T(X)]-A(\bgamma)+A(\bgamma_0),\\
				D(p_{\bgamma_0}\|p_{\bgamma})=(\bgamma_0-\bgamma)^\top \mathbb{E}_{0}[T(X)]-A(\bgamma_0)+A(\bgamma),
			\end{align*}
			which are twice continuously differentiable with respect to $\bgamma$ in $\Gamma$ based on Assumption (ii). Besides, we have
			\begin{align*}
			\nabla_{\bgamma}^2D(p_{\bgamma}\|p_{\bgamma_0})&\overset{(a)}{=}\nabla((\bgamma-\bgamma_0)^\top \nabla_{\bgamma}^2 A(\bgamma)),\\
			\nabla_{\bgamma}^2D(p_{\bgamma_0}\|p_{\bgamma})&=\nabla_{\bgamma}^2 A(\bgamma).
			\end{align*}
			where $(a)$ holds because $\mathbb{E}_{\bgamma}[T(X)]=\nabla_{\bgamma} A(\bgamma)$~\cite{wainwright2008graphical}.
			Based on Assumption (iii), $D(p_{\bgamma}\|p_{\bgamma_0})$ and $D(p_{\bgamma_0}\|p_{\bgamma})$ are strongly convex in $\bgamma$. Hence,
			the solutions to the minimizations are unique. Then we also have
			\begin{align*}
			\nabla_{\bgamma}D(p_{\bgamma}\|p_{\bgamma_0})=(\bgamma-\bgamma_0)^\top \nabla^2_{\bgamma}A(\bgamma),
			\end{align*}
			which means that $\nabla_{\bgamma}D(p_{\bgamma}\|p_{\bgamma_0})=0$ and $D(p_{\bgamma}\|p_{\bgamma_0})=0$ if and only if $\bgamma = \bgamma_0$. As $\bgamma_0\notin\Gamma$, $\min_{\bgamma\in\Gamma}D(p_{\bgamma}\|p_{\bgamma_0})>0$. Similarly, we have 
			\begin{align*}
			\nabla_{\bgamma}D(p_{\bgamma_0}\|p_{\bgamma})=-\nabla_{\bgamma}A(\bgamma_0)+\nabla_{\bgamma}A(\bgamma).
			\end{align*} 
			As $\nabla_{\bgamma}^2 A(\bgamma)$ assumed to be positive definite per Assumption (iii), then $\nabla_{\bgamma}A(\bgamma_0)=\nabla_{\bgamma}A(\bgamma)$ if and only if $\bgamma= \bgamma_0$. As $\bgamma_0\notin\Gamma$,
			we have $\min_{\bgamma\in\Gamma}D(p_{\bgamma_0}\|p_{\bgamma})>0$.

			For Condition (A3), we have
			\begin{align*}
				\xi(\bgamma)=(\bgamma-\bgamma_0)^\top T(X)-A(\bgamma)+A(\bgamma_0).
			\end{align*}
			Then $\mathbb{E}[\max_{\bgamma}|\xi(\bgamma)|^2]<\infty$ due to Assumptions (i) and (ii). 
			Let $\mathbf{e}$ be the all ones vector. For all $t>0$ and $t\be+\bgamma_0\in\Theta$, we have
			\begin{align*}
				&P_0\bigg(\max_{\bgamma\in\Gamma}|\nabla_{\bgamma}\xi(\bgamma)|>x\bigg)\\
				&=P_0\bigg(\max_{\bgamma\in\Gamma}\big|T(X)-\nabla_{\bgamma} A(\bgamma)\big|>x\bigg)\\
				&\overset{(a)}{=}P_0\bigg(\max_{\bgamma\in\Gamma}\big|T(X)-\mathbb{E}_{\bgamma}[T(X)]\big|>x\bigg)\\
				&\leq P_0\!\bigg(\!\big|T\!(X)\!-\!\mathbb{E}_{0}[T\!(X)]\big|\!+\max_{\bgamma\in\Gamma}\big|\mathbb{E}_0[T(X)]\!-\!\mathbb{E}_{\bgamma}[T(X)]\big|\!>\!x\!\bigg)\\
				&=P_0\!\bigg(\!\big|T(X)\!-\!\mathbb{E}_{0}[T\!(X)]\big|\!>\!x\!-\!\max_{\bgamma\in\Gamma}\!\big|\mathbb{E}_0[T(X)]\!-\!\mathbb{E}_{\bgamma}[T(X)]\big|\!\bigg)\\
				&\overset{(b)}{\leq} \exp\bigg(-tx+t\max_{\bgamma\in\Gamma}\big|\mathbb{E}_0[T(X)]-\mathbb{E}_{\bgamma}[T(X)]\big|\\
				&\quad+A(t\be+\bgamma_0)-A(\bgamma_0)-t|\nabla_{\bgamma} A(\bgamma)|_{\bgamma=\bgamma_0}|+\log2\bigg),
			\end{align*}
			where $(a)$ is based on the property $\mathbb{E}_{\bgamma}[T(X)]=\nabla_{\bgamma} A(\bgamma)$, $(b)$ is based on Markov's inequality and the fact that $\bbE_0[\exp(\langle t\be, T(X)-\bbE_0[T(X)]\rangle)]=\exp(A(t\be+\bgamma_0)-A(\bgamma_0)-t|\nabla_{\bgamma} A(\bgamma)|_{\bgamma=\bgamma_0}|)$~\cite{allan}. Denote $\tilde{x}=t\max_{\bgamma\in\Gamma}\big|\mathbb{E}_0[T(X)]-\mathbb{E}_{\bgamma}[T(X)]\big|+A(t\be+\bgamma_0)-A(\bgamma_0)-t|\nabla_{\bgamma} A(\bgamma)|_{\bgamma=\bgamma_0}|+\log2$. Then there exists $\alpha>1$ such that when $x>x_0=\max\{x_1,1\}$ (where $x_1$ is the solution to $tx_1-\tilde{x}=(\log x_1)^{\alpha}$ if it exists, else $x_1=0$),
			\begin{align*}
				P_0\bigg(\max_{\bgamma\in\Gamma}|\nabla_{\bgamma}\xi(\bgamma)|>x\bigg)\leq e^{-(tx-\tilde{x})}\leq e^{-(\log x)^{\alpha}},
			\end{align*}
			which shows that~\eqref{eqn:con} holds.}
			\end{example}

	\begin{example}[Gaussian distributions]  {\em 
	 For Gaussian distributions, $\bgamma=[\mu/\sigma^2,-1/2\sigma^2]^\top$, $T(x)=[x,x^2]^\top$, $A(\bgamma)=-\frac{\gamma_1^2}{4\gamma_2}-\frac{1}{2}\log(-2\gamma_2)$ and $h(x)=\frac{1}{\sqrt{2\pi}}$, where $\gamma_1$ and $\gamma_2$ are the elements of $\bgamma$. 	We consider the test  
	 \begin{align*}
	 	&H_0: \calN(0,1),\;\quad\bgamma_0=[0,-1/2]^T;\\
	 	&H_1: \calN(\mu,\sigma^2), \quad\bgamma=[\mu/\sigma^2,-1/2\sigma^2]^T\in\Gamma, \bgamma_0\notin\Gamma.
	 \end{align*}
	 We assume that $\Gamma$ is a convex and compact set and $\sigma^2>\frac{4\mu^2+1}{3\mu+1}$.

		 For this example, Assumption (i) (i.e., Condition (A1)) holds as we assume that  $\Gamma$ is a convex and compact set. Besides, $A(\bgamma)$ is thrice continuously differentiable and 
		\begin{align*}
		\frac{\partial A^2(\bgamma_{1})}{\partial \bgamma_{1}^2}=\left[
		\begin{matrix}
		& -\frac{1}{2\gamma_2} & \frac{\gamma_1}{2\gamma_2^2}\\
		&\frac{\gamma_1}{2\gamma_2^2} & -\frac{\gamma_1^2}{2\gamma_2^3}+\frac{1}{2\gamma_2^2}
		\end{matrix}
		\right],
		\end{align*}
		which is positive definite. Besides,
		\begin{align*}
		\frac{\partial ((\bgamma-\bgamma_0)^TA''(\bgamma))}{\partial \bgamma}=\left[
		\begin{matrix}
		&\frac{1}{4\gamma_2^2} &-\frac{\gamma_1}{2\gamma_2^3}\\
		&\frac{-\gamma_1}{2\gamma_2^3} &\frac{3\gamma_1}{4\gamma_2^4}-\frac{1}{2\gamma_2^3}-\frac{1}{\gamma_2^2}
		\end{matrix}\right],
		\end{align*}
		which is positive definite when $\sigma^2>\frac{4\mu^2+1}{3\mu+1}$. Thus, Assumptions (ii) and (iii) hold, which implies Condition (A2) holds. For Condition (A3), we have 
			\begin{align*}
			t\max_{\bgamma\in\Gamma}&\big|\mathbb{E}_0[T(X)]-\mathbb{E}_{\bgamma}[T(X)]\big|+A(t\be+\bgamma_0)-A(\bgamma_0)\\
			&-t|A'(\bgamma_0)|=t\max_{\bgamma\in\Gamma}\bigg|-\frac{\gamma_1^2}{4\gamma_2}-\frac{1}{2}\log(-2\gamma_2)\bigg|\\
			&-\frac{t^2}{4(t-1/2)}-\frac{1}{2}\log(-2(t-1/2))-t.
			\end{align*}
			Then we choose $t=\frac{1}{4}$, we have
				\begin{align*}
				\frac{1}{4}\max_{\bgamma\in\Gamma}&\big|\mathbb{E}_0[T(X)]-\mathbb{E}_{\bgamma}[T(X)]\big|+A\Big(\frac{1}{4}\be+\bgamma_0\Big)-A(\bgamma_0)\\
				&-\frac{1}{4}|A'(\bgamma_0)|=\frac{1}{4}\max_{\bgamma\in\Gamma}\bigg|-\frac{\gamma_1^2}{4\gamma_2}-\frac{1}{2}\log(-2\gamma_2)\bigg|+\frac{5}{16}.
				\end{align*}
			Denote $\tilde{x}=\frac{1}{4}\max_{\bgamma\in\Gamma}\Big|-\frac{\gamma_1^2}{4\gamma_2}-\frac{1}{2}\log(-2\gamma_2)\Big|-\frac{3}{16}+\frac{3}{2}\log 2$. Then there exists $\alpha>1$ such that when $x>x_0=\max\{x_1,1\}$ (where $x_1$ is the solution to $\frac{1}{4}x_1-\tilde{x}=(\log x_1)^{\alpha}$ if it exists, else $x_1=0$),
			\begin{align*}
				P_0\bigg(\max_{\bgamma\in\Gamma}|\nabla_{\bgamma}\xi(\bgamma)|>x\bigg)\leq e^{-(tx-\tilde{x})}\leq e^{-(\log x)^{\alpha}},
			\end{align*}
			which shows that Condition (A3) holds.	 }
	\end{example}

	Our first main result is Theorem~\ref{thm:firstorder} which characterizes the set of first-order error exponents under the probabilistic constraints on the stopping time in \eqref{eqn:probcons}.
	\begin{theorem}
		\label{thm:firstorder}
		For fixed $0<\epsilon<1$ and if Conditions (A1)--(A3) are satisfied, the set of $\epsilon$-achievable pair of error exponents is
		$$\calE_{\epsilon}(p_{0},\Gamma)=\left\{(E_0,E_1):
		\begin{aligned}
		&E_0\leq \min_{\gamma\in\Gamma} D(p_{\gamma}\|p_0),\\
		&E_1\leq \min_{\gamma\in\Gamma}D(p_0\|p_{\gamma}).
		\end{aligned}\right\}
		$$
		Furthermore, the corner point of this set is achieved by an appropriately defined sequence of GSPRTs.
	\end{theorem}
	 
		Theorem~\ref{thm:firstorder} shows that the $\epsilon$-achievable error exponent region is a rectangle.   
		In addition, Theorem~\ref{thm:firstorder} shows a strong converse result because the region does not depend on  the permissible error probability $0<\epsilon<1$.
 
	\subsection{Proof of Achievability of Theorem~\ref{thm:firstorder}} 
	\label{subsec:ach1}
	Let $\veps_0$ and $\veps_1$ be two positive numbers such that 
	$\veps_0 \in \big(0,\min_{\gamma\in\Gamma}D(p_{\gamma}\|p_0)\big)$ and $\veps_1\in \big(0,\min_{\gamma\in\Gamma}D(p_0\|p_{\gamma})\big)$. Let $(\delta_n,\tau_n)$ be the GSPRT with the thresholds $A_n:=n(\min_{\gamma\in\Gamma}D(p_{\gamma}\|p_0)-\veps_0)$ and $B_n:=n(\min_{\gamma\in\Gamma}D(p_0\|p_{\gamma})-\veps_1)$. Since Conditions (A1)--(A3) are satisfied, then from~\cite[Theorem~2.1]{Xiaoou} we have that
	\begin{align}
	 \liminf_{n\to\infty} \frac{1}{n}\log \frac{1}{ P_{0} (S_{\tau_n}\!>\!A_n)} &\!\geq\!\min_{\gamma\in\Gamma}D(p_{\gamma}\|p_0)\!-\!\veps_0\label{type11},\\
	 \liminf_{n\to\infty}  \frac{1}{n}\log\frac{1}{\displaystyle\sup_{\gamma\in\Gamma} P_{{\gamma}} (S_{\tau_n}\!<\!-B_n)}&\!\geq\! \min_{\gamma\in\Gamma} D(p_0\|p_{\gamma})\!-\!\veps_1\label{type22}.
	\end{align}
	
	Next we prove that the two probabilistic constraints in~\eqref{eqn:probcons} are satisfied for the GSPRT $(\delta_n,\tau_n)$ with thresholds $A_n$ and $B_n$. We first introduce the uniform weak law of large numbers (UWLLN)~\cite[Theorem 6.10]{bierens_2004}. 
	\begin{lemma}
	\label{lemm:uwlln}
		Let $\{X_j\}_{j=1}^{\infty}$ be a sequence of i.i.d.\ random vectors, and let $\gamma\in\Gamma$ be a nonrandom vector lying in a compact subset $\Gamma\subset\mathbb{R}^d$. Moreover, let $g(x,\gamma)$ be a Borel-measurable function on $\mathcal{X}\times \Gamma$ such that for each $x, g(x,\gamma)$ is continuous on~$\Gamma$. Finally, assume that $\mathbb{E}\left[\max_{\gamma\in\Gamma}|g(X_j,\gamma)|\right]<\infty$. Then for any $\delta>0$,
		\begin{align*}
		\lim_{n\to\infty}\mathbb{P}\left(\max_{\gamma\in\Gamma} \bigg|\frac{1}{n}\sum_{j=1}^n g(X_j,\gamma)-\mathbb{E}[g(X,\gamma)]\bigg|\geq\delta\right)=0.
		\end{align*}
	\end{lemma}

	Let $\tau':=\inf\{k:S_k<-B_n\}$. We observe that $\tau'\geq \tau_n$, so we have
	\begin{align*}
	P_{0}(\tau_n>n)\leq P_{0}(\tau'>n)=P_{0}\left(\max_{\gamma\in\Gamma}S_n(\gamma)\geq -B_n\right).
	\end{align*}
	Because
	\begin{align*}
	&\max_{\gamma\in\Gamma}\sum_{i=1}^n\log\frac{{p_{\gamma}}(X_i)}{{p_0}(X_i)}+n\min_{\gamma\in\Gamma} D(p_0\|p_{\gamma}) \\
	&\leq\max_{\gamma\in\Gamma}\left(\sum_{i=1}^n\log\frac{{p_{\gamma}}(X_i)}{{p_0}(X_i)}-n\mathbb{E}_{0} \left[\log\frac{{p_{\gamma}}(X)}{{p_0}(X)}\right]\right),
	\end{align*}
	and $\max_x f(x) - \min_x g(x)\le \max_x (f(x)-g(x)) \le \max_x |f(x)-g(x)|$, we have
	\begin{align*}
	&P_{0}\left(\max_{\gamma\in\Gamma}S_n(\gamma)\geq -n\Big(\min_{\gamma\in\Gamma}D(p_0\|p_{\gamma})-\veps_1\Big)\right)\\
	&\leq P_{0}\left(\max_{\gamma\in\Gamma}\bigg|\sum_{i=1}^n\log\frac{{p_{\gamma}}(X_i)}{{p_0}(X_i)}-n\mathbb{E}_{0} \left[\log\frac{{p_{\gamma}}(X)}{{p_0}(X)}\right]\bigg|\geq n\veps_1\right).
	\end{align*}
	Then by UWLLN, for $0<\epsilon<1$, there exists an $n_0(\epsilon)$, such that when $n>n_0(\epsilon)$, 
$$P_{0}\left(\max_{\gamma\in\Gamma}\bigg|\sum_{i=1}^n\log\frac{{p_{\gamma}}(X_i)}{{p_0}(X_i)} -  n\mathbb{E}_{0} \left[\log\frac{{p_{\gamma}}(X)}{{p_0}(X)}\right]\bigg|\ge n\veps_1\right)<\epsilon.$$
Therefore, $P_{0}(\tau_n>n)\leq P_{0}(\tau'>n)< \epsilon.$

	We now prove that $\sup_{\gamma\in\Gamma}P_{{\gamma}}(\tau_n>n)< \epsilon$. Define $\tau'':=\inf\{k:\max_{\gamma\in\Gamma}S_k(\gamma)> A_n\}$. We also have $\tau''\geq\tau_n$. Then for each $\gamma_0\in\Gamma$ and $t<0$, we have
	\begin{align*}
	P_{{\gamma_0}}(\tau_n>n)&\leq P_{{\gamma_0}}(\tau''>n)\\*
	&\leq P_{{\gamma_0}}\left(\max_{\gamma\in\Gamma}S_n(\gamma)\leq A_n\right)\\*
	&\leq P_{{\gamma_0}}\big(S_n(\gamma_0)\leq n(D(p_{\gamma_0}\|p_0)-\veps_0)\big)\\*
 	&\leq \frac{\mathrm{Var}(\xi(\gamma_0))}{n\eta_0^2}
 	\end{align*}
 	where the last step follows from Chebyshev's inequality~\cite{durrett2004probability}. Then based on Condition (A3) that $\mathbb{E}[\max_{\gamma}|\xi(\gamma)|^2]<\infty$ and $\eta_0$ does not depend on $\gamma_0$, there exists an $n_1(\epsilon)$ such that when $n>n_1(\epsilon)$, $\sup_{\gamma\in\Gamma}P_{{\gamma}}(\tau_n>n)< \epsilon.$ We have shown that when $n>\max\{n_0(\epsilon),n_1(\epsilon)\}$, the two probabilistic constraints~\eqref{eqn:probcons} are satisfied. Then together with~(\ref{type11}), (\ref{type22}) and the arbitrariness of $\veps_0$ and $\veps_1$, we show that any exponent pair $(E_0,E_1)$ such that $E_0\le \min_{\gamma\in\Gamma}D(p_{\gamma}\|p_0)$ and  $E_1\le \min_{\gamma\in\Gamma}D(p_0\|p_{\gamma})$ is in $\calE_{\epsilon}(p_{0},\Gamma)$

	\subsection{Proof of Strong Converse of Theorem~\ref{thm:firstorder}} 
	
The following lemma is taken from Li and Tan~\cite{yonglong}.
	\begin{lemma}
		\label{lemma:converse}
		Let $(\delta,\tau)$ be a sequential hypothesis test such that $P_{0}(\tau<\infty)=1$ and $\sup_{\gamma\in\Gamma}P_{{\gamma}}(\tau<\infty)=1$. 
		Then for any  event $F\in\calF_{\tau}$, $\lambda>0$ and for each $\gamma_0\in\Gamma$ we have 
		\begin{align*}
		P_{0}(F)-\lambda P_{{\gamma_0}}(F) &\leq P_{0}(S_{\tau}(\gamma_0)\leq-\log\lambda),\\
		P_{{\gamma_0}}(F)-\frac{1}{\lambda} P_{0}(F)&\leq P_{{\gamma_0}}(S_{\tau}(\gamma_0)\geq-\log\lambda).
		\end{align*}
	\end{lemma}	
	 Then we use Lemma~\ref{lemma:converse} to prove the converse part. 
	Let $(E_0,E_1)\in \calE_{\epsilon}(p_{0},\Gamma)$ such that $\min\{E_0,E_1\}>0$. Without loss of generality and by passing to a subsequence if necessary, we assume that there exists a sequence of sequential hypothesis tests $\{(\delta_n,\tau_n)\}_{n=1}^{\infty}$  such that $P_0(\tau_n<\infty)=1$ and $\sup_{\gamma\in\Gamma}P_{\gamma}(\tau_n<\infty)=1$ and 
	\begin{align}
	E_0&=\lim_{n\to\infty}\frac{1}{n}\log\frac{1}{\rvP_{1|0}(\delta_n,\tau_n)}, \label{exp1}\\*
	E_1&=\lim_{n\to\infty}\frac{1}{n}\log\frac{1}{\rvP_{0|1}(\delta_n,\tau_n)}. \nn
	\end{align}
	Let $Z_i(\tau_n)=\{\delta_n=i\}$ for $i=0,1$. Then $\rvP_{1|0}(\delta_n,\tau_n)=P_{0}(Z_1(\tau_n))$ and $\rvP_{0|1}(\delta_n,\tau_n)=\sup_{\gamma\in\Gamma}P_{{\gamma}}(Z_0(\tau_n))$. Using Lemma~\ref{lemma:converse} with the event $F=Z_0(\tau_n)$,  for each $\gamma_0\in\Gamma$ we have that 
	\begin{align*}
	&1-P_{0}(Z_1(\tau_n))-\lambda P_{{\gamma_0}}(Z_0(\tau_n))\\
	&\leq P_{0}(S_{\tau_n}(\gamma_0)\leq-\log\lambda)\\*
	&\leq P_{0}(S_{\tau_n}(\gamma_0)\leq-\log\lambda,\tau_n\leq n)+P_{0}(\tau_n>n),
	\end{align*}
	which further implies that
	\begin{align}
	\label{eqn:type II}
	&\log P_{{\gamma_0}}(Z_0(\tau_n)) \notag\\
	&\geq \log\bigg[ \frac{1}{\lambda}\Big(1-P_{0}(Z_1(\tau_n))-P_{0}(\tau_n>n)\notag\\
	&\qquad-P_{0}(S_{\tau_n}(\gamma_0)\leq-\log\lambda,\tau_n\leq n)\Big)\bigg].
	\end{align}
Similarly, for each $\gamma_0\in\Gamma$, we have that 
	\begin{align*}
	1-P_{{\gamma_0}}(&Z_0(\tau_n))-\frac{1}{\lambda} P_{0}(Z_1(\tau_n))\\
	&\leq P_{{\gamma_0}}(S_{\tau}(\gamma_0)\geq-\log\lambda,\tau_n\leq n)+P_{{\gamma_0}}(\tau_n>n),
	\end{align*}
		and when we set $E=Z_1(\tau_n)$, we have
	\begin{align}
	\label{eqn:type I}
	&\log P_{0}(Z_1(\tau_n)) \notag\\
	&\geq \log\Big[ \lambda\Big(1-P_{{\gamma_0}}(Z_0(\tau_n))-P_{{\gamma_0}}(\tau_n>n)\notag\\
	&\quad-P_{{\gamma_0}}(S_{\tau_n}(\gamma_0)\geq-\log\lambda,\tau_n\leq n)\Big)\Big].
	\end{align}
	
 Let $\delta$ be an arbitrary positive number and let $\log\lambda=n\left(D(p_0\|p_{\gamma_0})+\delta\right)$. We first bound the term
 \begin{align*}
 	&P_{0}(S_{\tau_n}(\gamma_0)\leq-\log\lambda,\tau_n\leq n)\\
 	&\leq P_{0}\left(\max_{1\leq k\leq n} \sum_{i=1}^k\log\frac{{p_0}(X_i)}{{p_{\gamma_0}}(X_i)}\geq n(D(p_0\|p_{\gamma_0})+\delta)\right).
 \end{align*}
 We note that $\big\{\log\frac{{p_0}(X_i)}{{p_{\gamma_0}}(X_i)}-D(p_0\|p_{\gamma_0})\big\}_{i=1}^n$ is an i.i.d. sequence. Besides, we have that $\bbE_0\left[-\xi(\gamma_0)-D(p_0\|p_{\gamma_0})\right]=0$ and $\mathrm{Var}\left(\xi(\gamma_0)\right)$ is finite based on Condition (A3). Then based on Kolmogorov's maximal inequality~\cite[Theorem 2.5.5]{durrett2004probability}, we have that
 \begin{align}
 P_{0}\bigg(\max_{1\leq k\leq n} \sum_{i=1}^k\log\frac{{p_0}(X_i)}{{p_{\gamma_0}}(X_i)}&\geq n(D(p_0\|p_{\gamma_0})+\delta)\bigg)\notag\\
 &\leq \frac{\mathrm{Var}\left(\xi(\gamma_0)\right)}{n\delta^2}.
 \end{align}
 Note that here we use the Kolmogorov's maximal inequality and it only requires that the  second moment of the log-likelihood ratio  is finite; this is a weaker condition than assuming that the   third absolute moment of the log-likelihood ratio is finite as in~\cite{yonglong}.
Then we have that
	 \begin{align}\label{im3}
	 \lim_{n\to\infty} \sup_{\gamma_{0} \in \Gamma}P_{0}\Bigg(\max_{1\leq k\leq n}\sum_{i=1}^k\log\frac{{p_0}(X_i)}{{p_{\gamma_0}}(X_i)}\geq \log\lambda\Bigg)=0.
	 \end{align}

When we set $-\log\lambda=n(D(p_{\gamma_0}\|p_0)+\delta)$ in (\ref{eqn:type I}), using similar arguments as in the derivation of~(\ref{im3}), we obtain
\begin{align} 
\lim_{n\to\infty}\sup_{\gamma_{0}\in \Gamma}P_{{\gamma_0}}\bigg(\max_{1\leq k\leq n}S_{k}(\gamma_0)\geq-\log\lambda\bigg)=0.\nn
\end{align}
From~\eqref{eqn:type II} and the fact that $P_{\gamma_0}(\tau_n>n)<\epsilon$, we have that
\begin{align*}
-\frac{1}{n} & \sup_{\gamma\in\Gamma}\log P_{\gamma}(Z_0(\tau_n))\\
&\leq \min_{\gamma\in\Gamma}(D(p_0\|p_{\gamma})+\delta)-\frac{1}{n}\log\Bigg(1-\rvP_{1|0}(\delta_n,\tau_n)-\epsilon\\
&\quad-\sup_{\gamma\in\Gamma}P_{0}\bigg(\max_{1\leq k\leq n} \sum_{i=1}^k\log\frac{{p_0}(X_i)}{{p_{\gamma}}(X_i)}\geq \log\lambda\bigg)\Bigg).
\end{align*}
From~(\ref{exp1}) it follows that $\lim_{n\to\infty}\rvP_{1|0}(\delta_n,\tau_n)=0$, which together with~(\ref{im3}) implies that
\begin{align*}
E_1= \lim_{n\to\infty}-\frac{1}{n}\log\rvP_{0|1}(\delta_n,\tau_n)\leq \min_{\gamma\in\Gamma}D(p_0\|p_{\gamma})+\delta.
\end{align*}
Similarly, we also obtain
\begin{align*}
E_0= \lim_{n\to\infty}-\frac{1}{n}\log \rvP_{1|0}(\delta_n,\tau_n)&\leq  \min_{\gamma\in\Gamma}D(p_{\gamma}\|p_0)+\delta.
\end{align*}
Due to the arbitrariness of $\delta$, letting $\delta\to 0^{+}$, we have that $E_0\le  \min_{\gamma\in\Gamma}D(p_{\gamma}\|p_0)$ and $E_1\leq \min_{\gamma\in\Gamma}D(p_0\|p_{\gamma})$, completing the proof of the strong converse as desired.

	\section{Second-order Asymptotics}\label{second-order}
	In the previous section, we considered the (first-order) error exponents of the sequential composite hypothesis testing problem under probabilistic constraints. While the result (Theorem~\ref{thm:firstorder}) is conclusive, there is often substantial motivation~\cite{tan2014asymptotic} to consider higher-order asymptotics due to finite-length considerations. To wit, the probabilistic bound observation length of the sequence~$n$ might be short and thus the exponents derived in the previous section will be overly optimistic. In this section, we quantify the backoff from the optimal first-order exponents by examining the second-order asymptotics. To do so, we make a set of somewhat more stringent conditions on the distributions and the uncertainty set $\Gamma$.  We first assume that the alphabet of the observations is the finite set $\calX=\{1,2,\dots,d\}$. Let $\mathcal{P}_{\mathcal{X}}$ be the set of probability mass functions with alphabet $\mathcal{X}$. In other words, $\calP_{\calX}$ is the probability simplex given by 
	$$\mathcal{P}_{\mathcal{X}}\!:=\!\bigg\{\!(q(1),\!q(2),\ldots,\!q(d))\!:\! \sum_{i=1}^{d}\!q(i)=1, \!q(i)\ge 0,~\forall\, i\!\in\!\calX\!\bigg\}.$$
	Similarly, define the open probability simplex 	$$\mathcal{P}_{\mathcal{X}}^+\!:=\!\bigg\{\!(q(1),\!q(2),\ldots,\!q(d))\!:\! \sum_{i=1}^{d}\!q(i)=1, \!q(i)> 0,~\forall\, i\!\in\!\calX\!\bigg\}.$$

	 Under hypothesis $H_0$, the underlying probability mass function is given by $\{p_0(i)\}_{i=1}^{d}$ and we assume that $p_0(i)>0$ for all $i\in\calX$. Under hypothesis $H_1$, the underlying probability mass function belongs to the set $\Gamma\subset \mathcal{P}_{\mathcal{X}}$. For any $\tilde{q}\in\mathcal{P}_{\mathcal{X}}$ and positive constant $\eta$, let $\calB(\tilde{q},\eta):=\{q \in\calP_{\calX}:|q(i)-\tilde{q}(i)|<\eta,~\forall \, i\in\calX\}$ be the open $\eta$-neighborhood of the point $\tilde{q}$. Let $\bgamma'$ be such that $D(p_{0}\|p_{\bgamma'})=\min_{\bgamma\in\Gamma}D(p_{0}\|p_{\bgamma})$. See Fig.~\ref{fig:linear} for an illustration of this projection.

\subsection{Other Assumptions and Preliminary Results}	 
\label{sec:assumptions}
	 We assume that $\Gamma$, which contains distributions supported on $\calX$,  satisfies the following conditions:
	\begin{enumerate}
	\item[(A1')] The set $\Gamma$ is equal to $\{\bgamma=\{\gamma_{i}\}_{i=1}^{d}:F(\bgamma)\le 0\}\in\calP_{\calX}$ for some piece-wise smooth convex function $F: \mathcal{P}_{\mathcal{X}}\to\mathbb{R}$.
	\item[(A2')] There exists a fixed constant $c_{0}>0$ such that $\min_{i\in\mathcal{X}}\gamma_{i}\ge c_{0}$ for all $\bgamma\in \Gamma$.
	\item[(A3')] The function $F$ is smooth (infinitely differentiable) on $\calB(\bgamma',\eta)$ for some $\eta>0$.
	\end{enumerate}

	The key tool used in the derivation of the second-order terms is a central limit-type result for $\max_{\bgamma\in\Gamma} \sum_{k=1}^n\log\frac{p_{\bgamma}(X_k)}{p_0(X_k)}$, the maximum of log-likelihood ratios of the observations over $\Gamma$. To simplify this quantity, we define the {\em empirical distribution} or {\em type}~\cite[Chapter~11]{Cover} of $X^n$ as $Q(i;X^n)=\sum_{k=1}^n\mathbbm{1}\{X_k=i\}/n$, for $i=1,2,\dots,d$. In the following, for the sake of brevity, we often suppress the dependence on the sequence $X^n$ and write $Q(i)$ in place of $Q(i;X^n)$, but we note that $Q$ is a {\em random} distribution induced by the observations $X^n$. Since $\calX$ is a finite set, we have 
	\begin{align}
	\label{eqn:Sn}
	S_n&=\max_{\bgamma\in\Gamma} \sum_{k=1}^n\log\frac{p_{\bgamma}(X_k)}{p_0(X_k)} \notag\\
	&=\max_{\bgamma\in\Gamma}\sum_{i=1}^d \sum_{k=1}^n \mathbbm{1}\{X_k=i\}\log\frac{\gamma_i}{p_0(i)}\notag\\
	&=n\max_{\bgamma\in\Gamma}\sum_{i=1}^d Q(i)\log\frac{\gamma_i}{p_0(i)}.
	\end{align} 
	The key in obtaining the central limit-type result for the sequence of random variables $\{S_n/\sqrt{n}\}_{n\in\bbN}$ is to solve the optimization problem in~(\ref{eqn:Sn}), or more precisely, to understand the properties of the optimizer to~\eqref{eqn:Sn}. Now we study the following optimization problem for a generic $q\in \mathcal{P}_{\mathcal{X}}$:
	\begin{align}
	\label{eqn:opt}
	\min_{\bgamma}\;\;& \sum_{i=1}^d q(i)\log\frac{p_0(i)}{\gamma_i}\notag\\*
	\mathrm{s.t.}\;\; & \sum_{i=1}^d \gamma_i=1,\\*
	&F(\bgamma)\leq 0\notag.
	\end{align}
	Let $\tilde{\bgamma}$ be an optimizer to the optimization problem~(\ref{eqn:opt}). The properties of  $\tilde{\bgamma}$  are provided in the following three lemmas.
	\begin{lemma}\label{lemma:solution}
	If  $q\in\mathcal{P}^{+}_{\mathcal{X}}$ and $q\not\in\Gamma$, then the optimizer $\tilde{\bgamma}$ is unique. 	\end{lemma}
	The existence and uniqueness of the optimizer of  the optimization problem~\eqref{eqn:opt}  follows from the strictly convexity of the function $\bgamma\mapsto\sum_{i=1}^d q(i)\log\frac{p_0(i)}{\gamma_i}$ on the compact convex (uncertainty) set $\Gamma$.
	
As the optimizer $\tilde{\bgamma}$ is unique, we can define the function
\begin{align*}
\bg(q)=(g_{1}(q),\ldots,g_{d}(q)) =:\tilde{\bgamma}.
\end{align*}
For the sake of convenience in what follows, define
\begin{align}\label{def:objfun}
f(q):=\sum_{i=1}^d q(i)\log\frac{p_0(i)}{g_{i}(q)}.
\end{align} 
Some key properties of $\bg(q)$ are provided in Lemma~\ref{lemma:continuity} and Lemma~\ref{lemma:property} in the Appendix.
By the definition of $\bgamma'$, it follows that $\bg(p_0)=\bgamma'$.  Without loss of generality, we assume 
$$
	\frac{\partial F(\bgamma')}{\partial\gamma_1}-\sum_{i=1}^d \gamma'_i\frac{\partial F(\bgamma')}{\partial\gamma'_i}\neq0.$$
	Then there exists $0<\bar{\eta}<\hat{\eta}$ such that for $q\in\mathcal{B}(p_{0},\bar{\eta})$, the following equation holds
	$$
	\frac{\partial F(\tilde{\bgamma})}{\partial\tilde{\gamma}_1}-\sum_{i=1}^d \tilde{\gamma}_i\frac{\partial F(\tilde{\bgamma})}{\partial\tilde{\gamma}_i}\neq0.$$
	Then for $q\in\mathcal{B}(p_{0},\bar{\eta})$,  the Jacobian of $( q(2), \ldots, q(d))$ with respect to $(\tilde{\gamma}_2,\ldots, \tilde{\gamma}_{d})$  is
	\begin{align*}
	\bJ(q)=\begin{bmatrix}
	&\frac{\partial q(2)}{\partial \tilde{\gamma}_2} &\frac{\partial q(2)}{\partial \tilde{\gamma}_3} &\cdots &\frac{\partial q(2)}{\partial \tilde{\gamma}_d}\vspace{.5em}\\
	&\frac{\partial q(3)}{\partial \tilde{\gamma}_2} &\frac{\partial q(3)}{\partial \tilde{\gamma}_3} &\cdots &\frac{\partial q(3)}{\partial \tilde{\gamma}_d}\vspace{.5em}\\
	&\vdots&\vdots&\ddots&\vdots\vspace{.5em}\\
	&\frac{\partial q(d)}{\partial \tilde{\gamma}_2} &\frac{\partial q(d)}{\partial \tilde{\gamma}_3} &\cdots &\frac{\partial q(d)}{\partial \tilde{\gamma}_d}
	\end{bmatrix}  \in\bbR^{(d-1)\times(d-1)}.
	\end{align*}
We now introduce the following regularity condition on the function $F$ at the point $p_{0}$:
\begin{itemize}
\item[(A4')] The Jacobian matrix $\bJ(p_{0})$ is of full rank (i.e., $\mathrm{rank}(\bJ(p_{0}))=d-1$).
\end{itemize}

\begin{figure*}[t]
\centering
\begin{overpic}[width=.84\textwidth]{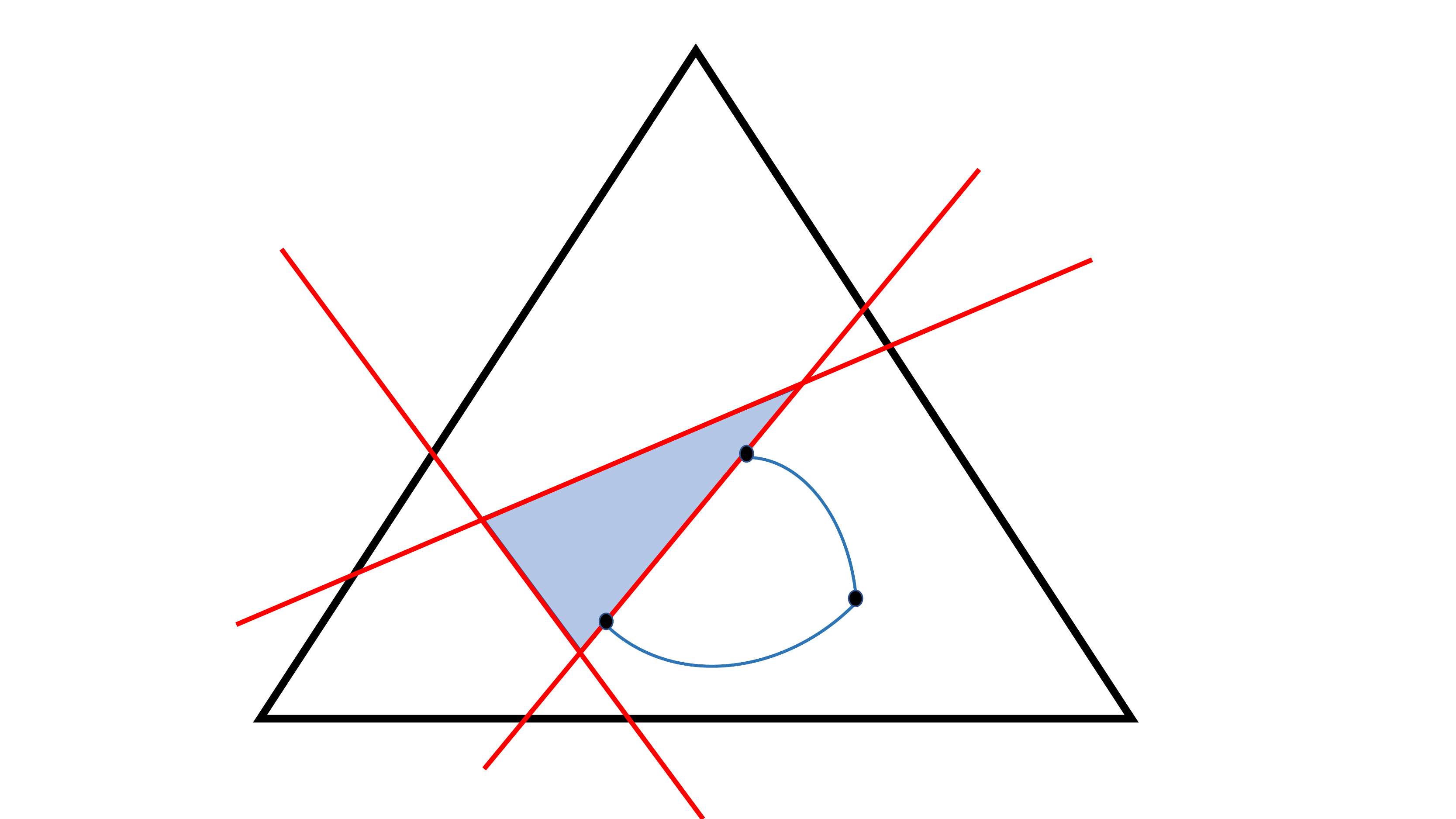}
\put(70,9){$\calP_\calX$} 
\put(68,45){$F_1$} 
\put(76,38){$F_2$} 
\put(15,38){$F_3$} 
\put(60,14){$p_0$} 
\put(41,20){$\Gamma$} 
\put(58,20){$D( p_0\|p_{\bgamma'}\!)$} 
\put(48.5,25){$\bgamma'$} 
\put(45,8.5){$D(p_{{\bgamma}^{*}}\|p_{0})$} 
\put(40,15){${\bgamma}^{*}$} 
\put(79,6){$(0,1,0)$} 
\put(5,6){$(1,0,0)$} 
\put(49,53){$(0,0,1)$} 
\end{overpic}
\caption{The set $\Gamma$ formed by the intersection of three halfspaces defined by $F_1, F_2$, and $F_3$. See Example~\ref{ex:linear}. }
\label{fig:linear}
\end{figure*}
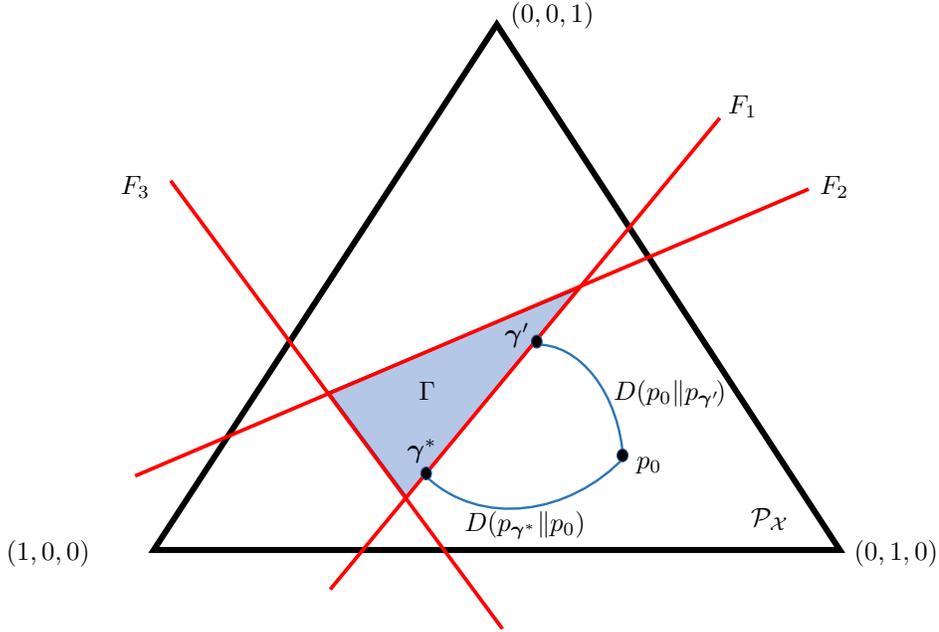

One may wonder whether the new assumptions we have stated are overly restrictive. In fact, they are not and there exist  interesting families of uncertainty sets that satisfy Assumptions (A1')--(A4'). A canonical  example of an uncertainty set $\Gamma$ that satisfies these conditions  is when~$F$ is piece-wise linear on the set $\calP_{\calX}^+$. Thus, $\Gamma$ is similar to a linear family~\cite{amari}, an important class of statistical models. 
	\begin{example} 
	\label{ex:linear}
	Let $\{F_{k}\}_{k=1}^{l}$ be a set of $l$ linear functions with domain $\mathbb{R}^{d}$ and let $\{\xi_{k}\}_{k=1}^{l}$ be a set of $l$ real numbers. Let $\Gamma=\bigcap_{k=1}^{l}\{(y_{1},\ldots,y_{d}):F_{k}(y_{1},\ldots,y_{d})\le \xi_{k}\}$. Assume $\{F_{k}\}_{k=1}^{l}$ and $\{\xi_{k}\}_{k=1}^{l}$ satisfy the following three conditions:
	\begin{itemize}
	\item The set $\Gamma\subset \calP_{\calX}^+$ and $F_{k}(p_{0})>\xi_{k}$ for some $k$;
	\item The minimizer $\bgamma'=\argmin_{\bgamma\in\Gamma} D(p_{0}\|p_{\bgamma})$ is such that $F_{1}(\bgamma')=\xi_{1}$ and $F_{k}(\bgamma')<\xi_{k}$ for $k\not=1$;
	\item Let $F_1 (y_{1},\ldots,y_{d})= \sum_{i=1}^d w_iy_i$ for some real coefficients $w_1, \ldots, w_d$.  One of the coefficients of $F_{1}$, i.e., one of the numbers in the set $\{w_i\}_{i=1}^d$,   is not equal to $\xi_{1}$.
	\end{itemize}
		\end{example}
			Intuitively, $\Gamma$ defined as the intersection of halfspaces (linear inequality constraints) is a polyhedron contained in the relative interior of $\mathcal{P}_{\mathcal{X}}$.   Fig.~\ref{fig:linear} provides an illustration for the ternary case $\mathcal{X}=\{1,2,3\}$.

	\begin{proposition}\label{lemma:linearfamily}
	The set $\Gamma$ described in Example~\ref{ex:linear} satisfies Conditions (A1')--(A4').
	\end{proposition}
	The proof of Proposition~\ref{lemma:linearfamily} is provided in   Appendix~\ref{app:linear}. 
	
	Now we are ready to state the promised central limit-type result for $S_{n}$, defined in \eqref{eqn:Sn}. Define the {\em relative entropy variance}~\cite{tan2014asymptotic} $$V(p\|q):=\mathrm{Var}_{p}\left[\log\frac{p(X)}{q(X)}\right]$$ and the Gaussian cumulative distribution function $\Phi(y) := \int_{-\infty}^y\frac{1}{\sqrt{2\pi}}e^{-u^2/2}\, \mathrm{d}u$. Then we have
 \begin{proposition}
		\label{lem:gaussian}
		Under Conditions (A1')--(A4'), if $\{X_{i}\}_{i=1}^{\infty}$ is a sequence of i.i.d.\ random variables with $P(X_{1}=i)=p_{0}(i)$ for all $i\in\mathcal{X}$, then $\{S_n\}_{n=1}^\infty$, defined in \eqref{eqn:Sn}, satisfies
		\begin{align*} 
		\sqrt{n}\bigg(\frac{S_{n}}{n}-D(p_0\|p_{\bgamma'})\bigg)\stackrel{\mathrm{d}}{\longrightarrow} \calN\big(0,V(p_0\|p_{\bgamma'})\big).
		\end{align*}
	\end{proposition}
	The proof of Proposition~\ref{lem:gaussian} can be found in Appendix~\ref{proof:gaussian}. 
	
	A major result in the statistics literature that bears some semblance to Proposition~\ref{lem:gaussian} is known as Wilks' theorem (see~\cite[Chapter~16]{vaart_1998} for example). For the case in which the null hypothesis is simple,\footnote{Wilks' theorem also applies to the case in which both the null and alternative hypotheses are composite, but we are only concerned with the simpler setting here.}  Wilks' theorem states that if the sequence of random variables $\{X_i\}_{i=1}^{\infty}$ is independently drawn from $p_0$ (the distribution of the null hypothesis), then (two times) the log-likelihood ratio statistic
	\begin{align*}
	2\max_{\bgamma\in\Gamma\cup \{p_0\}}\sum_{k=1}^n\log\frac{p_{\bgamma}(X_k)}{p_0(X_k)}\stackrel{\mathrm{d}}{\longrightarrow} \chi^2_{d-1},
	\end{align*}
	where $\chi_{d-1}^2$ is the chi-squared distribution with $d-1$ degrees of freedom. This result differs from Proposition~\ref{lem:gaussian} because in $S_n$ the maximization is taken over $\Gamma$ whereas in Wilks' theorem, it is taken over $\Gamma\cup\{p_0\}$. This results in different normalizations in the statements on convergence in distributions; in Proposition~\ref{lem:gaussian}, $S_n$ is normalized by $\sqrt{n}$ but there is no normalization of the log-likelihood ratio statistic in Wilks' theorem. This is because, for the former (our result), the dominant term is the first-order term in the Taylor expansion, but in the latter (Wilks' theorem), the dominant term is the second-order term.

	\begin{proposition}
	\label{lem:condition}
	Conditions (A1')--(A4') imply Conditions (A1)--(A3) in Section~\ref{first-order}.
	\end{proposition}
	The proof of Proposition~\ref{lem:condition} is provided in Appendix~\ref{proof:condition}. Thus, we see that the assumptions used to derive the first-order results are less restrictive than those for the second-order result that we are going to state in the next subsection.

	\subsection{Definition and Main Results}
	We say that a second-order exponent pair $(G_0,G_1)$ is {\em $\epsilon$-achievable under the probabilistic constraints} if there exists a sequence of sequential hypothesis tests $\{(\delta_n,\tau_n)\}_{n=1}^{\infty}$ that satisfies the probabilistic constraints on the stopping time in~\eqref{eqn:probcons} and 
	\begin{align*}
	&G_0\leq \liminf_{n\to\infty} \frac{1}{\sqrt{n}}\bigg(\log\frac{1}{\rvP_{1|0}(\delta_n,\tau_n)}-nD(p_{\bgamma^{*}}\|p_0)\bigg),\\*
	&G_1\leq \liminf_{n\to\infty} \frac{1}{\sqrt{n}}\bigg(\log\frac{1}{\rvP_{0|1}(\delta_n,\tau_n)}-nD(p_0\|p_{\bgamma'})\bigg),
	\end{align*}
where $\bgamma^{*}=\argmin_{\bgamma\in\Gamma} D(p_{\bgamma}\|p_0)$, which is unique (see Proposition~\ref{lem:condition} which implies that Condition (A2) is satisfied). The set of all $\epsilon$-achievable second-order exponent pairs $(G_0,G_1)$ is denoted as $\mathcal{G}_{\epsilon}(p_{0},\Gamma)$, the {\em second-order error exponent region}. 
	The set of second-order error exponents $\mathcal{G}_{\epsilon}(p_0,\Gamma)$ is stated in the following theorem.
	\begin{theorem}
		\label{thm:second-order}
		If Conditions (A1')--(A4') are satisfied, for any $0<\epsilon<1$, the second-order error exponent region is
		\begin{align*}
		\mathcal{G}_{\epsilon}(p_{0},\Gamma)=\left\{(G_0,G_1) \in\bbR^2:\! 
		\begin{array}{rl}
		 G_0&\hspace{-.11in}\leq \Phi^{-1}(\epsilon)\sqrt{V(p_{\bgamma^{*}}\|p_0)}\\ 
		 G_1&\hspace{-.11in}\leq \Phi^{-1}(\epsilon)\sqrt{V(p_0\|p_{\bgamma'})}
		\end{array}\!
		\right\}.
		\end{align*}
		Furthermore, the boundary of this set is achieved by an appropriately defined sequence of GSPRTs.
	\end{theorem}
This theorem states that the backoffs from the (first-order) error exponents are of orders $\Theta(1/\sqrt{n})$ and the implied constants are given by $\Phi^{-1}(\epsilon)\sqrt{V(p_{\bgamma^{*}}\|p_0)}$ and $ \Phi^{-1}(\epsilon)\sqrt{V(p_0\|p_{\bgamma'})}$. Thus, we have stated a set of sufficient conditions on the distributions and the uncertainty set $\Gamma$ (namely (A1')--(A4'))  for which the second-order terms are analogous to that for simple sequential hypothesis testing under the probabilistic constraints derived by Li and Tan~\cite{yonglong}. However, the techniques used to derive Theorem~\ref{thm:second-order} are more involved compared to those for the probabilistic constraints in \cite{yonglong}. This is because we have to derive the asymptotic distribution of the {\em maximum} of a set of log-likelihood ratio terms (cf.\ Proposition~\ref{lem:gaussian}). This constitutes our main contribution in this part of the paper.
\subsection{Proof of the Achievability Part of Theorem~\ref{thm:second-order}}
	\label{subsec:ach}
	
The proof of achievability consists of two parts. We first prove the desired upper bound on type-I error probability and the maximal type-II error probability under an appropriately defined sequence of  GSPRTs. Then we prove that the probabilistic constraints are satisfied.

We start with the proof of the first part. Let $\eta_0$ and $\eta_1$ be such that $\eta_0,\eta_1\in(0,\epsilon)$.  Let $(\delta_n,\tau_n)$ be the GSPRT with thresholds $$A_n:=n\min_{\bgamma\in\Gamma}\bigg(D(p_{\bgamma}\|p_0)+\Phi^{-1}(\epsilon-\eta_0)\sqrt{\frac{V(p_{\bgamma}\|p_0)}{n}}\bigg),$$ and $$B_n:=n\min_{\bgamma\in\Gamma}\bigg(D(p_0\|p_{\bgamma})+\Phi^{-1}(\epsilon-\eta_1)\sqrt{\frac{V(p_0\|p_{\bgamma})}{n}}\bigg).$$  

Based on Proposition~\ref{lem:condition}, we know that (A1)--(A3) are satisfied. Hence, from~\cite[Theorem~2.1]{Xiaoou} we have that
 \begin{align*}
 P_{0} (S_{\tau_n}>A_n)
 \leq e^{-A_n}\quad\mbox{and}\quad\sup_{\bgamma\in\Gamma} P_{{\bgamma}} (S_{\tau_n}<-B_n)\leq e^{-B_n}.
 \end{align*}
 
	To simplify $A_n$ and $B_n$, we introduce an approximation lemma from~\cite[Lemma 48]{polyanskiy2010channel}. 
	\begin{lemma}
		\label{ref:poly}
		Let $\Gamma$ be a compact metric space. Suppose $h:\Gamma\to\mathbb{R}$ and $k:\Gamma\to \mathbb{R}$ are continuous, then we have
		\begin{align*}
		\max_{\bgamma\in\Gamma}[nh(\bgamma)+\sqrt{n}k(\bgamma)]=n h^{*}+\sqrt{n} k^{*} +o(\sqrt{n}),
		\end{align*}
		where $h^{*}:=\max_{\bgamma\in\Gamma}h(\bgamma)$ and $k^{*}:=\sup_{\bgamma:h(\bgamma)=h^{*}}k(\bgamma)$.
	\end{lemma}
Here we take $h(\bm{\gamma})=-D(p_0\|p_{\bgamma})$ and $k(\bm{\gamma})=-\Phi^{-1}(\epsilon-\eta_1)\sqrt{V(p_0\|p_{\bgamma})}$. Based on Lemma~\ref{ref:poly} and the fact that $\bgamma\mapsto D(p_0\|p_{\bgamma})$ has a unique minimizer $\gamma'$ (see Assumption (A2) which is implied by Proposition~\ref{lem:condition}), we have
 \begin{align}
 \label{eqn:sec_ach1}
 &\min_{\bgamma\in\Gamma}\left(nD(p_0\|p_{\bgamma})+\sqrt{{nV(p_0\|p_{\bgamma})}}\Phi^{-1}(\epsilon-\eta_1)\right)\notag \\
 &=nD(p_0\|p_{\bgamma'}) +{\Phi^{-1}(\epsilon-\eta_1)}\sqrt{nV(p_0\|p_{\bgamma'})}+o(\sqrt{n}).
 \end{align}
 Similarly, we have
 \begin{align}
 \label{eqn:sec_ach2}
 &\min_{\bgamma\in\Gamma}\left(nD(p_{\bgamma}\|p_0)+\sqrt{{nV(p_{\bgamma}\|p_0)}}\Phi^{-1}(\epsilon-\eta_0)\right)\notag\\
 &= nD(p_{\bgamma^{*}}\|p_0) +{\Phi^{-1}(\epsilon-\eta_0)}\sqrt{nV(p_{\bgamma^{*}}\|p_0)}+o(\sqrt{n}).
 \end{align}
 
	Thus, based on~\eqref{eqn:sec_ach1} and~\eqref{eqn:sec_ach2}, the arbitrariness of $\eta_0$ and $\eta_1$ and the continuity of $\Phi^{-1}$, we obtain
	\begin{align}
	\label{eqn:second1}
	\!\liminf_{n\to\infty} \frac{1}{\sqrt{n}}\Big( \log &\frac{1}{P_{0} (S_{\tau_n}>A_n)}-{n}D(p_{\bgamma^{*}}\|p_0)\Big)\notag\\ &\geq\Phi^{-1}(\epsilon)\sqrt{{V(p_{\bgamma^{*}}\|p_0)}},
	\end{align}
	and
	\begin{align}
	\label{eqn:second2}
	 \liminf_{n\to\infty}  \frac{1}{\sqrt{n}}\Big( \log&\frac{1}{\sup_{\gamma\in\Gamma} P_{{\gamma}} (S_{\tau_n}<-B_n)}-n D(p_0\|p_{\bgamma'})\Big)\notag\\ &\geq\Phi^{-1}(\epsilon)\sqrt{{V(p_0\|p_{\bgamma'})}}.
	\end{align}

	
	Next we prove that the probabilistic constraints for the sequence of GSPRTs $\{(\delta_n,\tau_n)\}_{n=1}^{\infty}$ are satisfied. Let $\tau':=\inf\{k:\max_{\bgamma\in\Gamma}S_k(\gamma)<-B_n\}$. We observe that $\tau'\geq \tau_n$ with probability 1.
	Thus, we have
	\begin{align}
	&P_{0}(\tau_n>n)\notag\\
	&\leq P_{0}(\tau'>n)\notag\\
	&\le P_{0}\left(\max_{\bgamma\in\Gamma}S_n(\bgamma)\geq -B_n\right)\notag\\
	&=P_0\bigg(\min_{\bgamma\in\Gamma}n\sum_{i=1}^d Q(i)\log\frac{p_0(i)}{\gamma_i}\leq B_n\bigg)\notag\\
	&\leq P_0\bigg(\min_{\bgamma\in\Gamma}\sqrt{n}\Big(\sum_{i=1}^d Q(i)\log\frac{p_0(i)}{\gamma_i}- D(p_0\|p_{\bgamma'})\Big)\notag\\
	&\quad\leq  {\Phi^{-1}(\epsilon-\eta_1)}\sqrt{V(p_0\|p_{\bgamma'})}\bigg)\notag\\*
	& \to \epsilon-\eta_1 \label{eqn:eps_eta1}\\
	&<\epsilon, \label{eqn:seccons1}
	\end{align}
	where \eqref{eqn:eps_eta1} is from Proposition~\ref{lem:gaussian}. Hence, $P_{0}(\tau_n>n)< \epsilon$ for sufficiently large $n$.

We now prove that $\sup_{\bgamma\in\Gamma}P_{\gamma}(\tau_n>n)< \epsilon$. Let $\tau'':=\inf\{k:\max_{\bgamma\in\Gamma}S_k(\bgamma)> A_n\}$. We also have $\tau''\geq\tau_n$ with probability 1. Then by the Berry-Esseen Theorem~\cite{10.2307/1990053}, for any $\bgamma_0\in\Gamma$, we have
\begin{align}
\label{eqn:seccon2}
&P_{{\bgamma_0}}(\tau_n>n)\notag\\
&\leq P_{{\bgamma_0}}(\tau''>n)\notag \\*
&\leq P_{{\bgamma_0}}\left(\max_{\bgamma\in\Gamma}S_n(\bgamma)\leq A_n\right)\notag\\*
&\leq \!P_{{\bgamma_0}}\!\bigg(S_n(\bgamma_0)\!\leq \!n\Big(D(p_{\bgamma_0}\!\|p_0)\!+\!\sqrt{\frac{V\!(p_{\bgamma_0}\|p_0)}{n}}\Phi^{-1}(\epsilon\!-\eta_0)\!\Big)\!\bigg)\notag\\*
&\leq \epsilon-\eta_0+\frac{T_1}{\sqrt{n}},
\end{align}
where $T_1$ is a positive finite constant depending only on $\mathrm{Var}_{\bgamma_0}(\xi(\bgamma_0))$ and $\mathbb{E}_{\bgamma_0}[|\xi(\bgamma_0)|^3]$. As stated in Condition (A2') (i.e., that $\gamma_i\geq c_0>0, i=1,\dots,d$) and $p_0(i)>0, i=1,\dots,d$, thus $\mathbb{E}_{\bgamma}[|\xi(\bgamma)|^3]$ is uniformly bounded on $\Gamma$. Then for every $0<\epsilon<1$, there exists an integer $n_1(\epsilon)$ which does not depend on $\bgamma$, such that when $n>n_1(\epsilon)$, $P_{{\bgamma_0}}(\tau>n)\leq \epsilon-\eta_0/2<\epsilon$.
Since $\bgamma_0\in\Gamma$ is arbitrary, $\sup_{\bgamma\in\Gamma}P_{{\bgamma}}(\tau>n)< \epsilon$.

We have shown that the two probabilistic constraints~\eqref{eqn:seccons1} and~\eqref{eqn:seccon2} are satisfied for sufficiently large $n$. Then together with~\eqref{eqn:second1} and~\eqref{eqn:second2}, we have shown that any second-order error exponent pair $(G_0,G_1)$ such that $G_0\geq\Phi^{-1}(\epsilon)\sqrt{V(p_{\bgamma^{*}}\|p_0)}$ and $G_1\geq{\Phi^{-1}(\epsilon)}\sqrt{V(p_0\|p_{\bgamma'})}$ belongs to $\calG_{\epsilon}(p_0,\Gamma)$.

\subsection{Proof of the Converse Part of Theorem~\ref{thm:second-order}} 
For each $\bgamma_0\in\Gamma$, from~\cite{yonglong}, we know that 
\begin{align*}
-\frac{1}{\sqrt{n}}&\log P_{\bgamma_0}(Z_0(\tau_n))\\
&\leq\sqrt{n}D(p_0\|p_{\bgamma_0})+\sqrt{V(p_0\|p_{\bgamma_0})}\Phi^{-1}(\epsilon)+\alpha_n,
\end{align*}
where $\alpha_n\to 0$ as $n\to\infty$.
Now we want to find the optimal upper bound for all $\gamma\in\Gamma$, which means we need to obtain
\begin{align*}
-\frac{1}{\sqrt{n}}&\sup_{\bgamma\in\Gamma}\log P_{\bgamma}(Z_0(\tau_n))\\
&\leq\min_{\bgamma\in\Gamma}\bigg(\sqrt{n}D(p_0\|p_{\bgamma})+\sqrt{V(p_0\|p_{\bgamma})}\Phi^{-1}(\epsilon)+\alpha_n\bigg).
\end{align*}

Similar to the analysis in achievability part, we use Lemma~\ref{ref:poly} and obtain that
\begin{align*}
\limsup_{n\to\infty} \frac{1}{\sqrt{n}}\Big(\log& \frac{1}{\rvP_{1|0}(\delta_n,\tau_n)}-n D(p_0\|p_{\bgamma'})\Big)\\
&\leq \Phi^{-1}(\epsilon)\sqrt{V(p_0\|p_{\bgamma'})}.
\end{align*}
Similarly, we have that
\begin{align*}
\limsup_{n\to\infty} \frac{1}{\sqrt{n}}\Big( \log& \frac{1}{\rvP_{0|1}(\delta_n,\tau_n)}-n D(p_{\bgamma^{*}}\|p_0)\Big)\\
&\leq \Phi^{-1}(\epsilon)\sqrt{V(p_{\bgamma^{*}}\|p_0)},
\end{align*}	
which completes the proof of the converse.

\appendix
\renewcommand{\thesubsection}{\Alph{subsection}}
In the appendix, we provide some key properties of $\bg(q)$ in Lemmas~\ref{lemma:continuity} and \ref{lemma:property} and their proofs. We also present the proofs of Propositions \ref{lemma:linearfamily}, \ref{lem:gaussian}, and \ref{lem:condition}.
\subsection{Properties of  $\bg(q)$}
\begin{lemma}\label{lemma:continuity}
	If $q\in\mathcal{P}^{+}_{\mathcal{X}}$ and $q\not\in\Gamma$, then the following properties of the optimizer $\tilde{\bgamma}=\bg(q)$ hold.
	\begin{itemize}
		\item[(i)]\label{item:continuity} 	The function $\bg(q)$ is continuous on $\calP_{\calX}^+\setminus\Gamma$;
		\item[(ii)] \label{item:differentiable} There exists $\hat{\eta}>0$ such that for $q\in\mathcal{B}(p_{0},\hat{\eta})$, $F$ is smooth (infinitely differentiable) at $\tilde{\bgamma}$;
		\item[(iii)] For $q\in\mathcal{B}(p_{0},\hat{\eta})$, the optimizer $\tilde{\bgamma}$ is such that $F(\tilde{\bgamma})=0$ (i.e., $\tilde{\bgamma}$ is on the boundary of the uncertainty set);
		\item[(iv)] For $q\in\mathcal{B}(p_{0},\hat{\eta})$,
		there exists a symbol $j\in\calX$ such that 
		\begin{equation}
			\frac{\partial F(\tilde{\bgamma})}{\partial\gamma_j}-\sum_{i=1}^d\tilde{\gamma}_i\frac{\partial F(\tilde{\bgamma})}{\partial\gamma_i}\neq0; \label{eqn:symbolj}
		\end{equation}
		\item[(v)] For $q\in\mathcal{B}(p_{0},\hat{\eta})$, $i\in\mathcal{X}$ and $i\not=j$ ($j\in\calX$ is the symbol that satisfies~\eqref{eqn:symbolj} in Part (iv) above),
		\begin{align}
			\label{eqn:Q}
			q(i)&=\tilde{\gamma}_i+\frac{(q(j)-\tilde{\gamma}_j)\tilde{\gamma}_i}{\tilde{\gamma}_{j}\big(\frac{\partial F(\tilde{\bgamma})}{\partial\gamma_j}-\sum_{k=1}^d\tilde{\gamma}_k\frac{\partial F(\tilde{\bgamma})}{\partial\gamma_k}\big)}\notag\\
			&\qquad\times\bigg(\frac{\partial F(\tilde{\bgamma})}{\partial\gamma_i}-\sum_{k=1}^d\tilde{\gamma}_k\frac{\partial F(\tilde{\bgamma})}{\partial\gamma_k}\bigg).
		\end{align}
	\end{itemize}
\end{lemma}

\begin{IEEEproof}
We first prove Part (i) of Lemma~\ref{lemma:continuity}. Assume, to the contrary, that $\bg(q)$ is not continuous at some $q\in \mathcal{P}^{+}_{\mathcal{X}}\setminus\Gamma$. Then there exists a positive number~$\kappa$ and a sequence $\{q_{k}\}_{k=1}^{\infty}\subset \mathcal{P}^{+}_{\mathcal{X}}\setminus\Gamma$ such that $q_{k}\to q$ as $k\to\infty$ and $\sum_{i=1}^{d}| g_{i}(q_{k})-g_{i}(q)|\ge \kappa$ for all $k\in\mathbb{N}$. From the definition of $\bg(q_{k})$ and the fact that $p_{0}\in \mathcal{P}^{+}$, there exists $\hat{\kappa}>0$ such that
\begin{align}\label{eqn:cont1}
\sum_{i=1}^{d}q_{k}(i)\log\frac{p_{0}(i)}{g_{i}(q_{k})}<\sum_{i=1}^{d}q_{k}(i)\log\frac{p_{0}(i)}{g_{i}(q)}-\hat{\kappa},
\end{align}
for all $k\in\bbN$. 
From Condition (A2') and the fact that $\{\bg(q_{k})\}_{k=1}^{\infty}\subset\Gamma$, there exists a constant $M < \infty$ such that
$$\sup_{k\in\mathbb{N},i\in\mathcal{X}}\left|\log\frac{p_{0}(i)}{g_{i}(q_{k})}\right|\le M,$$
which further implies that
\begin{align}\label{eqn:cont2}
&\limsup_{k\to\infty}\sum_{i=1}^{d}q_{k}(i)\log\frac{p_{0}(i)}{g_{i}(q_{k})}\notag\\
&=\limsup_{k\to\infty}\bigg(\sum_{i=1}^{d}\big(q(i)+(q_{k}(i)-q(i))\big)\log\frac{p_{0}(i)}{g_{i}(q_{k})}\bigg)\notag\\
&=\limsup_{k\to\infty}\sum_{i=1}^{d}q(i)\log\frac{p_{0}(i)}{g_{i}(q_{k})}.
\end{align}
Combining~(\ref{eqn:cont1}) and~(\ref{eqn:cont2}), we have that
\begin{align*}
&\limsup_{k\to\infty}\sum_{i=1}^{d}q(i)\log\frac{p_{0}(i)}{g_{i}(q_{k})}\\
&\le \limsup_{k\to\infty}\sum_{i=1}^{d}q_{k}(i)\log\frac{p_{0}(i)}{g_{i}(q)}-\hat{\kappa}\\
&=\sum_{i=1}^{d}q(i)\log\frac{p_{0}(i)}{g_{i}(q)}-\hat{\kappa},
\end{align*}
which contradicts the fact that 
$$\sum_{i=1}^{d}q(i)\log\frac{p_{0}(i)}{g_{i}(q_{k})}\ge \sum_{i=1}^{d}q(i)\log\frac{p_{0}(i)}{g_{i}(q)}.$$
Hence $\bg(q)$ is continuous on $\mathcal{P}^{+}_{\mathcal{X}}\setminus\Gamma$.

We next prove Part (ii) of Lemma~\ref{lemma:continuity}. From the continuity of $\bg(q)$ (as proved above), there exists $\hat{\eta}>0$ such that 
$$\{\tilde{\bgamma}: \tilde{\bgamma}=\bg(q) \,\mbox{for some $q\in\mathcal{B}(p_0,\hat{\eta})$}\}\subset\mathcal{B}(\bg(p_0),\eta),$$ which, together with Condition (A3') implies Part (ii) of Lemma~\ref{lemma:continuity}.

We now proceed to prove Part (iii) of Lemma~\ref{lemma:continuity}. Recall that the optimizer $\tilde{\bgamma}$ is obtained from the optimization problem~\eqref{eqn:opt}. Its corresponding Lagrangian is 
\begin{align*}
L(\bgamma,\lambda,\mu)=\sum_{i=1}^d q(i)\log\frac{p_0(i)}{\gamma_i}+\lambda\bigg(\sum_{i=1}^d \gamma_i-1\bigg)+\mu F(\bgamma).
\end{align*}
For $q\in\mathcal{B}(\bg(q),\hat{\eta})$, $F(\bgamma)$ is smooth at $\tilde{\bgamma}$ (the previous part). Hence using the Karush–Kuhn–Tucker (KKT) conditions~\cite{boyd2004convex}, the optimizer $\tilde{\bgamma}$ satisfies the first-order stationary conditions, which are
\begin{align}\label{eqn:stationary}
-\frac{q(i)}{\tilde{\gamma}_i}+\lambda+\mu\frac{\partial F(\bgamma)}{\partial\gamma_i}\bigg|_{\bgamma=\tilde{\bgamma}}=0, \quad\forall\, i=1,\dots,d.
\end{align}
The complementary slackness condition is $\mu F(\tilde{\bgamma})=0$, which implies that either $\mu=0$ or $F(\tilde{\bgamma})=0$. When $\mu=0$, we have 
\begin{align*}
q(i)=\lambda \tilde{\gamma}_i\;\Longleftrightarrow\; \lambda=1\;\Longleftrightarrow\; \tilde{\gamma}_i=q(i),
\end{align*}
which is impossible as $q\notin\Gamma$. Thus, it holds that $F(\tilde{\bgamma})=0$, which means the optimizer lies on the boundary of the set $\Gamma$.

We then proceed to prove Part (iv) of Lemma~\ref{lemma:continuity}. If $$
	\frac{\partial F(\bgamma)}{\partial \gamma_j}\Big|_{\bgamma=\tilde{\bgamma}}-\sum_{i=1}^d\tilde{\gamma}_i\frac{\partial F({\bgamma})}{\partial\gamma_i}\Big|_{\bgamma=\tilde{\bgamma}}=0$$ for all $j\in\mathcal{X}$, then $\big\{\frac{\partial F(\bgamma)}{\partial \gamma_j}\big|_{\bgamma=\tilde{\bgamma}}\big\}_{j=1}^{d}$ are all equal. Combining  this fact with~(\ref{eqn:stationary}), we have that
	$q=\tilde{\bgamma}$, which contradicts the fact that $q\not\in \Gamma$.

Finally, we prove Part (v) of Lemma~\ref{lemma:continuity}. Combining the constraints in~\eqref{eqn:opt} and~\eqref{eqn:stationary}, we can obtain $q$ in terms of  $\lambda$  as
	\begin{align}\label{eqn:qj}
	q(j)=\tilde{\gamma}_j-\mu\tilde{\gamma}_j\sum_{i=1}^d \tilde{\gamma}_i\frac{\partial F(\bgamma)}{\partial\tilde{\gamma}_i}+\mu\tilde{\gamma}_j\frac{\partial F(\bgamma)}{\partial\tilde{\gamma}_j}
	\end{align}
	for all $j=1,2,\dots,d$.
Then we obtain $\mu$ in terms of $q(j)$ as:
\begin{align}\label{eqn:mu}
	\mu=\frac{1}{\tilde{\gamma}_{j}}\bigg(\frac{\partial F(\bgamma)}{\partial\tilde{\gamma}_j}-\sum_{i=1}^d \tilde{\gamma}_i\frac{\partial F(\bgamma)}{\partial\tilde{\gamma}_i}\bigg)^{-1}(q(j)-\tilde{\gamma}_j).
	\end{align}
	Then substituting~(\ref{eqn:mu}) into~(\ref{eqn:qj}), we have the desired formula. This completes the proof of Lemma \ref{lemma:continuity}.
\end{IEEEproof}
	\begin{lemma}\label{lemma:property}
	Let $\hat{\eta}$ be as given in Lemma~\ref{lemma:continuity}. Suppose  $q\in\mathcal{B}(p_{0},\hat{\eta})$,  and $\Gamma$ satisfies (A1')--(A4'). Then,
	\begin{enumerate}
		\item[(i)]\label{item:smoothness} The function $\bg(q)$ is smooth on $\calB(p_{0},\zeta)$ for some $\zeta>0$ and satisfies the following equality
		\begin{align}
			&\sum_{j=1}^d \frac{q(j)}{g_j(q)}\frac{\partial g_j(q)}{\partial q(i)}=0,\nn
		\end{align}
		for all $q\in\mathcal{B}(p_0,\zeta)$.
		\item[(ii)]\label{item:objectivefunction}  The function $f$, defined in \eqref{def:objfun}, is smooth on $\calB(p_{0},\zeta)$ and its first- and second-order derivatives are
		\begin{align}
			\frac{\partial f(q)}{\partial q(j)}&=\log\frac{p_{0}(j)}{g_{j}(q)}+\sum_{i=1}^{d}\frac{q(i)}{g_{i}(q)}\frac{\partial g_i(q)}{\partial q(j)},\label{eqn:firstd}\\
			\frac{\partial^2 f(q)}{\partial q(j)^2}
			&=-\frac{2}{g_j(q)}\frac{\partial g_j(q)}{\partial q(j)}
			-\sum_{i=1}^d\bigg[-\frac{q(i)}{g_i(q)^2}\notag\\&\qquad\times\bigg(\frac{\partial g_i(q)}{\partial q(j)}\bigg)^2+\frac{q(i)}{g_i(q)}\frac{\partial^2g_i(q)}{\partial q(j)^2} \bigg],\quad\mbox{and}\nn\\
			\frac{\partial^2 f(q)}{\partial q(j)\partial q(i)}&=-\frac{1}{g_j(q)}\frac{\partial g_j(q)}{\partial q(i)}-\frac{1}{g_i(q)}\frac{\partial g_i(q)}{\partial q(i)}\notag\\
			&\qquad-\sum_{l=1}^d\bigg[-\frac{q(l)}{g_l(q)^2}\frac{\partial g_l(q)}{\partial q(j)}\frac{\partial g_l(q)}{\partial q(i)}\notag\\
			&\qquad\quad+\frac{q(l)}{g_l(q)}\frac{\partial^2 g_l(q)}{\partial q(j)\partial q(i)} \bigg]\quad\mbox{for}\; i \ne j.
		\end{align}
	\end{enumerate}
\end{lemma}
\begin{IEEEproof}
Now we prove Part (i) of Lemma~\ref{lemma:property}. As $F(\bgamma)$ is smooth and $\bJ(p_{0})$ is of full rank, there exists $\zeta>0$ such that $\bJ(q)$ is of full rank for all $q\in \mathcal{B}(p_{0},\zeta)$. Then by the inverse function theorem~\cite[Theorem 2.11]{Spivak71}, $\tilde{\bgamma}=\bg(q)$ is differentiable in $q$. We multiply ${\partial g_j(q)}/{\partial q(i)}$ on both sides of~\eqref{eqn:stationary} and sum from $j=1$ to $d$  to obtain 
\begin{align}
\label{eqn:diff}
\sum_{j=1}^d\frac{q(j)}{g_j(q)}\frac{\partial g_j(q)}{\partial q(i)}=\lambda\sum_{j=1}^d\frac{\partial g_j(q)}{\partial q(i)}+\mu \sum_{j=1}^d\frac{\partial F(\bg(q))}{\partial g_j(q)}\frac{\partial g_j(q)}{\partial q(i)}.
\end{align}
We differentiate the first constraint $\sum_{i=1}^d \tilde{\gamma}_i=\sum_{i=1}^dg_i(q)=1$ with respect to $q$ on both sides to obtain
\begin{align}
\label{eqn:diff1}
\sum_{j=1}^d\frac{\partial g_j(q)}{\partial q(i)}=0.
\end{align}
From~Part (iii) of Lemma \ref{lemma:continuity} it follows that $F(\tilde{\bgamma})=F(\bg(q))=0$, which means that the function formed by the composition of $F$ and $\bg$ is always $0$ for all the $q\in\mathcal{B}(p_{0},\hat{\eta})$ . Therefore, the derivative of the composition of $F$ and $\bg$ with respect to $q$ is~$0$, i.e.,
\begin{align}
\label{eqn:diff2}
\frac{\partial F(\bg(q))}{\partial q(i)}=\sum_{j=1}^d\frac{\partial F(\bg(q))}{\partial g_j(q)}\frac{\partial g_j(q)}{\partial q(i)}=0.
\end{align}
Substituting~\eqref{eqn:diff1} and~\eqref{eqn:diff2} back into~\eqref{eqn:diff}, we have that
\begin{align*}
\sum_{j=1}^d\frac{q(j)}{g_j(q)}\frac{\partial g_j(q)}{\partial q(i)}=0,
\end{align*}
as desired.

Part (ii) of Lemma~\ref{lemma:property} follows from straightforward, albeit tedious, calculations. This completes the proof of Lemma \ref{lemma:property}.
\end{IEEEproof}
\subsection{Proof of Proposition~\ref{lemma:linearfamily}}
\label{app:linear}
Assume $F_1(\gamma)=F_1(\gamma_1,\ldots, \gamma_d)=\sum_{i=1}^d w_i\gamma_i$. Without loss of generality, we assume  $w_1\not=\xi_{1}$. 
Conditions (A1')--(A3') clearly hold. Hence from Part (ii) of Lemma~\ref{lemma:continuity} there exists $\hat{\eta}$ such that for all $q\in\mathcal{B}(p_{0},\hat{\eta})$, the optimizer $\tilde{\bgamma}$ of the optimization problem~\eqref{eqn:opt} is such that $F_{1}(\tilde{\bgamma})=\xi_{1}$ and that $F_{k}(\tilde{\bgamma})<\xi_{k}$ for all $k\not=1$. Note that ${\partial F_1(\bgamma)}/{\partial \gamma_i}=w_i$. Then for $q\in\mathcal{B}(p_{0},\hat{\eta})$, using the KKT conditions, we obtain the first-order optimality conditions for the optimizer $\tilde{\bgamma}$:
\begin{align}
\sum_{i=1}^d\tilde{\gamma}_i&=1,\nn\\
\sum_{i=1}^d w_i\tilde{\gamma}_i&=\xi_{1},\label{eqn:gamma2}\\
 q(i)&=\lambda_1\tilde{\gamma}_i+\lambda_2\tilde{\gamma}_iw_i.\nn
\end{align}
Hence,
\begin{align}
\label{eqn:lambda2}
\lambda_2=\frac{q(1)-\tilde{\gamma}_1}{\tilde{\gamma}_1(w_1-\xi_{1})}.
\end{align}
Substituting~\eqref{eqn:lambda2} into~\eqref{eqn:gamma2}, we obtain
\begin{align*} 
q(i)=\tilde{\gamma}_i\bigg(1+\frac{(q(1)-\tilde{\gamma}_1)(w_i-\xi_{1})}{\tilde{\gamma}_1(w_1-\xi_{1})}\bigg).
\end{align*}
Thus, the Jacobian  of $(q(2),\ldots, q(d))$ at $(\tilde{\gamma}_2, \ldots, \tilde{\gamma}_d)$ is the following $(d-1)\times (d-1)$ diagonal matrix:
\begin{align*}
\bJ(q)&=\mathrm{diag}\!\bigg[1\!+\!\frac{(q(1)\!-\!\tilde{\gamma}_1)(w_2\!-\!\xi_{1})}{\tilde{\gamma}_1(w_1-\xi_{1})},\!1\!+\!\frac{(q(1)\!-\!\tilde{\gamma}_1)(w_3\!-\!\xi_{1})}{\tilde{\gamma}_1(w_1-\xi_{1})},\\
&\qquad\ldots,1+\frac{(q(1)-\tilde{\gamma}_1)(w_d-\xi_{1})}{\tilde{\gamma}_1(w_1-\xi_{1})}\bigg].
\end{align*}
 Since $p_{0}(i)>0$ for all $i=1,2,\dots,d$, the diagonal terms in the Jacobian $\bJ(p_{0})$ are non-zero. Thus, $\mathrm{det}(\bJ(p_{0}))\neq 0$, which proves that Condition (A4') holds for the set $\Gamma$ in Example~\ref{ex:linear}.

\subsection{Proof of Proposition~\ref{lem:gaussian}}
\label{proof:gaussian}
We  now prove the promised central limit-type result for the sequence of random variables $\{S_n/\sqrt{n}\}_{n\in\bbN}$. 
Let $z\in(0,1)$. Let $\zeta$ be given as in Part (i) of Lemma~\ref{lemma:property} and define the {\em $\zeta$-typical set} \begin{align*}
&\calT_{\zeta}^{(n)}=\calT_{\zeta}^{(n)}(p_0)\\
&=\bigg\{x^n \in\calX^n:\Big|\Big( \frac{1}{n}\sum_{k=1}^n\mathbbm{1} \{ x_k=i\}\Big) -p_0(i) \Big|<\zeta,~\forall\, i\in\calX\bigg\}.
\end{align*}
This is the set of sequences whose types are near $p_0$.  The key idea is to  perform a Taylor expansion of the function $f(Q)=\sum_{i=1}^d Q(i)\log\frac{p_0(i)}{g_i(Q)}$  (defined in \eqref{def:objfun}) at the point $Q=p_0$ and analyze the asymptotics of the various terms in the expansion. For brevity, define the deviation of the type  $Q$ of $X^n$ from the true distribution at symbol $i\in\calX$ as 
$$
\Delta_i:=Q(i)-p_0(i).
$$ 
For $q\in \mathcal{B}(p_{0},\zeta)$, let $\bH(q) \in\bbR^{d\times d}$ be the Hessian matrix of $f(q)$. This is well defined because $f(\cdot)$ is twice continuously differentiable on $\mathcal{B}(p_{0},\zeta)$ according to Part (ii) of Lemma~\ref{lemma:property}.   If $x^{n}\in\calT_{\zeta}^{(n)}$, then $Q\in \mathcal{B}(p_{0},\zeta)$. 
Thus for $Q\in \mathcal{B}(p_{0},\zeta)$, using Taylor's theorem we have the expansion
\begin{align}
f(Q) &=f(p_{0})+(\nabla f(p_{0}))^\top(Q-p_0)\notag\\*
&\qquad+\frac{1}{2}({Q}-p_0)\bH(\tilde{Q})({Q}-p_0)^\top \nonumber\\
&=\sum_{i=1}^d p_0(i)\log \frac{p_0(i)}{g_i(p_0)}+\sum_{i=1}^d \log\frac{p_0(i)}{g_i(p_0)}\Delta_i\notag\\
&\qquad- \sum_{i=1}^d  \sum_{j=1}^d \frac{p_0(j)}{g_j(p_0)}\frac{\partial g_j(q)}{\partial q(i)}\bigg|_{q=p_0}\Delta_i\notag\\
&\qquad+\frac{1}{2}({Q}-p_0) \bH(\tilde{Q})( {Q}-p_0)^\top\label{eqn:taylor1}\\
&= D(p_0\|p_{\bgamma'})+\sum_{i=1}^d \log \frac{p_0(i)}{g_i(p_0)}\Delta_i\notag\\
&\qquad+\frac{1}{2}({Q}- p_0)\bH(\tilde{Q})({Q}-p_0)^\top\label{eqn:taylor2},
\end{align}
where $\tilde{Q}$ lies on the line segment between $Q$ and $p_0$,~(\ref{eqn:taylor1}) follows from~(\ref{eqn:firstd}) in Lemma~\ref{lemma:property} and~\eqref{eqn:taylor2} follows from Part~(i) of Lemma~\ref{lemma:property}.  Note that we represent probability mass functions as {\em row} vectors.

Then for $Q\in \mathcal{B}(p_{0},\zeta)$, from~(\ref{eqn:taylor2}), we have that
\begin{align}
\label{eqn:taylorapprox}
&\min_{\bgamma\in\Gamma}\bigg(\sqrt{n}\sum_{i=1}^d Q(i)\log\frac{p_0(i)}{\gamma_i}\bigg)-\sqrt{n}D(p_0\|p_{\bgamma'})\notag\\
&=\sqrt{n}\big(f(Q)-D(p_0\|p_{\bgamma'})\big)\notag\\
&=\sum_{i=1}^d \sqrt{n}\Delta_i\log\frac{p_0(i)}{g_i(p_0)}+\frac{\sqrt{n}}{2}({Q}-p_0)\bH(\tilde{Q})({Q}-p_0)^\top.
\end{align}

Let $\lambda_{\min}(\bH(q))$ and $\lambda_{\max}(\bH(q))$ be the smallest and largest eigenvalues of $\bH(q)$, respectively. From Part (i) of Lemma~\ref{lemma:property},   it follows that $f(\cdot)$ is smooth on $\mathcal{B}(q_{0},\zeta)$, which implies that there exists two constants $\tilde{c}$ and $\tilde{C}$ such that
\begin{align}\label{eqn:eigenvalue}
-\infty<\tilde{c}&<\min_{q\in \mathcal{B}(q_{0},\zeta)}\lambda_{\min}(\bH(q))\notag\\
&\le \max_{q\in \mathcal{B}(q_{0},\zeta)}\lambda_{\max}(\bH(q))<\tilde{C}<\infty.
\end{align}
Then we have the upper bound shown in~\eqref{eqn:ub} (at the top of the next page),
\begin{figure*}[t]
\begin{align}
P_0&\bigg(\min_{\bgamma\in\Gamma}\sqrt{n}\bigg(\sum_{i=1}^d Q(i)\log\frac{p_0(i)}{\gamma_i}-D(p_0\|p_{\bgamma'})\bigg)\leq\Phi^{-1}(z)\sqrt{V(p_0\|p_{\bgamma'})}\bigg)\notag\\
&\le P_0\bigg(\min_{\bgamma\in\Gamma}\sqrt{n}\bigg(\sum_{i=1}^d Q(i)\log\frac{p_0(i)}{\gamma_i}-D(p_0\|p_{\bgamma'})\bigg)\leq\Phi^{-1}(z)\sqrt{V(p_0\|p_{\bgamma'})},X^n\in\calT_{\zeta}^{(n)}\bigg)+P_0(X^n\notin\calT_{\zeta}^{(n)})\notag\\
&=P_0\bigg(\sum_{i=1}^d \sqrt{n}b_i\Delta_i
+\frac{\sqrt{n}}{2}({Q}-p_0)\bH(\tilde{Q})({Q}-p_0)^\top\leq\Phi^{-1}(z)\sqrt{V(p_0\|p_{\bgamma'})},X^n\in\calT_{\zeta}^{(n)}\bigg)+P_0(X^n\notin\calT_{\zeta}^{(n)})\label{eqn:upper1}\\
&\leq P_0\bigg(\sum_{i=1}^d \sqrt{n}\Delta_i
\log\frac{p_{0}(i)}{g_{i}(p_{0})}+\frac{\lambda_{\min}(\bH(\tilde{Q}))}{2}\sum_{i=1}^d \sqrt{n}\Delta_i^2\leq\Phi^{-1}(z)\sqrt{V(p_0\|p_{\bgamma'})},X^n\in\calT_{\zeta}^{(n)}\bigg)+P_0(X^n\notin\calT_{\zeta}^{(n)})\notag\\
&\le P_0\bigg(\sum_{i=1}^d \sqrt{n}\Delta_i
\log\frac{p_{0}(i)}{g_{i}(p_{0})}+\frac{\tilde{c}}{2}\sum_{i=1}^d \sqrt{n}\Delta_i^2\leq\Phi^{-1}(z)\sqrt{V(p_0\|p_{\bgamma'})},X^n\in\calT_{\zeta}^{(n)}\bigg)+P_0(X^n\notin\calT_{\zeta}^{(n)})\label{eqn:upper3}\\
&\le  P_0\bigg(\sum_{i=1}^d \sqrt{n}\Delta_i
\log\frac{p_{0}(i)}{g_{i}(p_{0})}+\frac{\tilde{c}}{2}\sum_{i=1}^d \sqrt{n}\Delta_i^2\leq\Phi^{-1}(z)\sqrt{V(p_0\|p_{\bgamma'})}\bigg)+d\exp(-2n\eta^2),\label{eqn:ub}
\end{align}\hrulefill
\end{figure*}
in which~(\ref{eqn:upper1}) follows from the fact that $Q\in\mathcal{B}(p_{0},\zeta)$ for $x^{n}\in\calT_{\zeta}^{(n)}$ and~(\ref{eqn:taylorapprox}),~(\ref{eqn:upper3}) follows from~(\ref{eqn:eigenvalue}), and~(\ref{eqn:ub}) holds by the union bound and Hoeffding's inequality.  Similarly, we can obtain the lower bound shown in~\eqref{eqn:lb} (also shown on  the next page).
\begin{figure*}
\begin{align}
P_0&\bigg(\min_{\bgamma\in\Gamma}\sqrt{n}\bigg(\sum_{i=1}^d Q(i)\log\frac{p_0(i)}{\gamma_i}-D(p_0\|p_{\bgamma'})\bigg)\leq\Phi^{-1}(z)\sqrt{V(p_0\|p_{\bgamma'})}\bigg)\notag\\
&\geq P_0\bigg(\sum_{i=1}^d \sqrt{n}b_i\Delta_i
+\frac{\sqrt{n}}{2}({Q}-p_0)\bH(\tilde{Q})({Q}-p_0)^\top\leq\Phi^{-1}(z)\sqrt{V(p_0\|p_{\bgamma'})},X^n\in\calT_{\zeta}^{(n)}\bigg)\notag\\
&\geq P_0\bigg(\sum_{i=1}^d \sqrt{n}b_i\Delta_i
+\frac{\lambda_{\max}(\bH(\tilde{Q}))}{2} \sum_{i=1}^d \sqrt{n}\Delta_i^2\leq\Phi^{-1}(z)\sqrt{V(p_0\|p_{\bgamma'})},X^n\in\calT_{\zeta}^{(n)}\bigg)\notag\\
&\ge P_0\bigg(\sum_{i=1}^d \sqrt{n}\Delta_i\log\frac{p_{0}(i)}{g_{i}(p_{0})}
+\frac{\tilde{C}}{2} \sum_{i=1}^d \sqrt{n}\Delta_i^2\leq\Phi^{-1}(z)\sqrt{V(p_0\|p_{\bgamma'})}, X^n\in\calT_{\zeta}^{(n)}\bigg)\notag\\
&\ge P_0\bigg(\sum_{i=1}^d \sqrt{n}\Delta_i\log\frac{p_{0}(i)}{g_{i}(p_{0})}
+\frac{\tilde{C}}{2} \sum_{i=1}^d \sqrt{n}\Delta_i^2\leq\Phi^{-1}(z)\sqrt{V(p_0\|p_{\bgamma'})}\bigg)-P_{0}\big(X^n\not\in\calT_{\zeta}^{(n)}\big)\notag\\
&\ge P_0\bigg(\sum_{i=1}^d \sqrt{n}\Delta_i\log\frac{p_{0}(i)}{g_{i}(p_{0})}
+\frac{\tilde{C}}{2} \sum_{i=1}^d \sqrt{n}\Delta_i^2\leq\Phi^{-1}(z)\sqrt{V(p_0\|p_{\bgamma'})}\bigg)-d\exp(-2n\zeta^2).\label{eqn:lb}
\end{align}\hrulefill
\end{figure*}
One can verify that
 \begin{align}\label{eqn:sumdeltai}
&n\sum_{i=1}^d \Delta_i\log\frac{p_{0}(i)}{g_{i}(p_{0})}\notag\\
&=\sum_{k=1}^n\bigg(\sum_{i=1}^d \big(\mathbbm{1}\{X_k=i\}-p_0(i)\big)\log\frac{p_{0}(i)}{g_{i}(p_{0})}\bigg)
\end{align}
 and the variance 
 \begin{align}
& \mathrm{Var}_0\bigg[\sum_{i=1}^d(\mathbbm{1}\{X_1=i\}-p_0(i))\log\frac{p_{0}(i)}{g_{i}(p_{0})}\bigg]\notag\\*
&=\mathbb{E}_0\bigg[\bigg(\sum_{i=1}^d(\mathbbm{1}\{X_1=i\}-p_0(i))\log\frac{p_{0}(i)}{g_{i}(p_{0})}\bigg)^2\bigg] \label{eqn:zero_mean} \\*
 &=\mathbb{E}_0\bigg[\sum_{i=1}^d\Big(\log\frac{p_{0}(i)}{g_{i}(p_{0})}\Big)^2(\mathbbm{1}\{X_1=i\}-p_0(i))^2 \notag\\* 
 &\qquad+ 2\sum_{j\neq i}  (\mathbbm{1}\{X_1=i\}\!-p_0(i))(\mathbbm{1}\{X_1=j\}\!-p_0(j)) \notag\\*
 &\qquad\quad\times\Big(\log\frac{p_{0}(i)}{g_{i}(p_{0})}\Big)\Big(\log\frac{p_{0}(j)}{g_{j}(p_{0})} \Big)\bigg]\notag\\*
 & = \sum_{i=1}^d(1-p_0(i))p_0(i)\log^{2}\frac{p_{0}(i)}{g_{i}(p_{0})}\notag\\*
 &\qquad -2\sum_{i\neq j}p_0(i)p_0(j)\Big(\log\frac{p_{0}(i)}{g_{i}(p_{0})}\Big)\Big(\log\frac{p_{0}(j)}{g_{j}(p_{0})}\Big)\label{eqn:simplify_variance}\\
 &=\sum_{i=1}^d p_0(i)\log^{2}\frac{p_{0}(i)}{g_{i}(p_{0})}-\sum_{i=1}^d p_0(i)^2\log^{2}\frac{p_{0}(i)}{g_{i}(p_{0})}\notag\\*
 &\qquad -2\sum_{i\neq j}p_0(i)p_0(j)\Big(\log\frac{p_{0}(i)}{g_{i}(p_{0})}\Big)\Big(\log\frac{p_{0}(j)}{g_{j}(p_{0})}\Big)\notag\\*
 &=V(p_0\|p_{\bgamma'}), \notag 
 \end{align}
where \eqref{eqn:zero_mean} follows from $$\mathbb{E}_0\bigg[\sum_{i=1}^d(\mathbbm{1}\{X_1=i\}-p_0(i))\log\frac{p_{0}(i)}{g_{i}(p_{0})}\bigg]=0,$$ and \eqref{eqn:simplify_variance} follows  from
\begin{align*}
&\sum_{i\neq j}\mathbb{E}_{0}\bigg[ (\mathbbm{1}\{X_1=i\}-p_0(i))(\mathbbm{1}\{X_1=j\}-p_0(j))\\
&\quad\times\Big(\log\frac{p_{0}(i)}{g_{i}(p_{0})}\Big)\Big(\log\frac{p_{0}(j)}{g_{j}(p_{0})}\Big) \bigg]\\
&=-\sum_{i\neq j}p_0(i)p_0(j)\Big(\log\frac{p_{0}(i)}{g_{i}(p_{0})}\Big)\Big(\log\frac{p_{0}(j)}{g_{j}(p_{0})}\Big)
\end{align*}
and \begin{align*}
&\mathbb{E}_0\bigg[\sum_{i=1}^d(\mathbbm{1}\{X_k=i\}-p_0(i))^2\log^{2}\frac{p_{0}(i)}{g_{i}(p_{0})}\bigg]\\
&=\sum_{i=1}^d(1-p_0(i))p_0(i)\log^{2}\frac{p_{0}(i)}{g_{i}(p_{0})}.
\end{align*}

 Therefore $n\sum_{i=1}^d \Delta_i\log\frac{p_{0}(i)}{g_{i}(p_{0})}$ is a sum of i.i.d.\ random variables $\big\{\sum_{i=1}^d(\mathbbm{1}\{X_k=i\}-p_0(i))\log\frac{p_{0}(i)}{g_{i}(p_{0})}\big\}_{k=1}^{n}$ with mean~$0$ and variance $V(p_0\|p_{\bgamma'})$. 
Hence, by the central limit theorem,
\begin{align*}
&\sum_{i=1}^d \sqrt{n}\Delta_i\log\frac{p_{0}(i)}{g_{i}(p_{0})}\stackrel{\mathrm{d}}{\longrightarrow} \mathcal{N}(0,V(p_0\|p_{\bgamma'})).
\end{align*}
Together with the fact that $\sum_{i=1}^d \sqrt{n}\Delta_i^2\to 0$ almost surely, this implies that
\begin{align}
&\sum_{i=1}^d \sqrt{n}\Delta_i\log\frac{p_{0}(i)}{g_{i}(p_{0})}+\frac{\tilde{c}}{2}\sum_{i=1}^d \sqrt{n}\Delta_i^2\stackrel{\mathrm{d}}{\longrightarrow}\mathcal{N}(0,V(p_0\|p_{\bgamma'})), \label{eqn:limit1}
\end{align} 
and
\begin{align}
&\sum_{i=1}^d \sqrt{n}\Delta_i\log\frac{p_{0}(i)}{g_{i}(p_{0})}+\frac{\tilde{C}}{2}\sum_{i=1}^d \sqrt{n}\Delta_i^2\stackrel{\mathrm{d}}{\longrightarrow}\mathcal{N}(0,V(p_0\|p_{\bgamma'})).\label{eqn:limit2}
\end{align}
Then combining~(\ref{eqn:ub}),~(\ref{eqn:lb}),~(\ref{eqn:limit1}) and~(\ref{eqn:limit2}), we have that 
\begin{align}
\limsup_{n\to\infty}&\;P_0\bigg(\min_{\bgamma\in\Gamma}\sqrt{n}\Big(\sum_{i=1}^d Q(i)\log\frac{p_0(i)}{\gamma_i}-D(p_0\|p_{\bgamma'})\Big)\notag\\
&\leq\Phi^{-1}(z)\sqrt{V(p_0\|p_{\bgamma'})}\bigg)\leq z,\label{eqn:gaussian1}
\end{align} 
and
\begin{align}
 \liminf_{n\to\infty}&\; P_0\bigg(\min_{\bgamma\in\Gamma}\sqrt{n}\Big(\sum_{i=1}^d Q(i)\log\frac{p_0(i)}{\gamma_i}-D(p_0\|p_{\bgamma'})\Big)\notag\\
 &\leq\Phi^{-1}(z)\sqrt{V(p_0\|p_{\bgamma'})}\bigg)\geq z.\label{eqn:gaussian2}
\end{align}
Since $z\in (0,1)$ is arbitrary,  it follows from~\eqref{eqn:gaussian1} and~\eqref{eqn:gaussian2} that 
\begin{align*}
\!\min_{\bgamma\in\Gamma}\!\sqrt{n}\!\bigg(\!\sum_{i=1}^d\! Q(i)\!\log\frac{p_0(i)}{\gamma_i}\!-\!D(p_0\|p_{\bgamma'})\!\bigg)\!\stackrel{\mathrm{d}}{\longrightarrow}\!\calN(0,V(p_0\|p_{\bgamma'})),
\end{align*}
which completes the proof of Proposition~\ref{lem:gaussian}.

 \subsection{Proof of Proposition~\ref{lem:condition}}
 \label{proof:condition}
 	We now show that Conditions (A1')--(A4') imply Conditions (A1)--(A3). Condition~(A1) is easily verified by Condition~(A1'). As $\calX=\{1,2,\dots,d\}$, we have
 	\begin{align*}
 	D(p_0\|p_{\bgamma})=\sum_{i=1}^d p_0(i)\log\frac{p_0(i)}{\gamma_i},
 	\end{align*}
 	and
 	\begin{align*}
 	D(p_{\bgamma}\|p_0)=\sum_{i=1}^d \gamma_i\log\frac{\gamma_i}{p_0(i)}.
 	\end{align*}
 	Combining Condition (A2') which says that $\min_{i=1,\ldots,d}\gamma_i\ge c_0>0$ for all $\gamma\in\Gamma$  and $\min_{i=1,\ldots,d}p_0(i)>0$, $D(p_0\|p_{\bgamma})$ and $D(p_{\bgamma}\|p_0)$ are uniformly bounded  and twice continuously differentiable on $\Gamma$. As $p_0\notin\Gamma$, $D(p_0\|p_{\bgamma})>0$ and $D(p_{\bgamma}\|p_0)> 0$, which together with the compactness of $\Gamma$, implies that 
	\begin{align}\label{eqn:minimization}
	\min_{\bgamma\in\Gamma}D(p_0\|p_{\bgamma})>0\quad\mbox{and}\quad\min_{\bgamma\in\Gamma}D(p_{\bgamma}\|p_0)>0.
	\end{align} From~\cite[Theorem 2.7.2]{Cover}, $D(p_0\|p_{\bgamma})$ is strictly convex in $(p_{0},\bgamma)$, which, together with the fact that $\Gamma$ is compact and convex, implies the uniqueness of the minimizers to the two optimization problems in~(\ref{eqn:minimization}).
	
 For Condition~(A3), as  $\calX$ is a finite alphabet and  Condition~(A2') holds, it can be easily checked that $\mathbb{E}[\max_{\bgamma\in\Gamma}|\xi(\bgamma)|^2]<\infty$. Note that $$\nabla_{\bgamma}\xi(\bgamma)= \Big(\frac{\mathbbm{1}\{X=1\}}{\gamma_1}, \ldots, \frac{\mathbbm{1}\{X=d\}}{\gamma_d} \Big)^\top.$$ We can define the finite number $x_0:=\max_{\bgamma\in\Gamma}\max_{i\in\calX}1/\gamma_i\le 1/c_0$ (because Condition~(A2') mandates that $\min_{i=1,\ldots,d}\gamma_i\ge c_0>0$  for all $\bgamma\in\Gamma)$. With this choice, trivially,  for all $x>x_0$,
 	\begin{align*}
 	P_{0}\Bigg(\max_{\bgamma\in\Gamma}|\nabla_{\bgamma}\xi(\bgamma)|>x\Bigg)=0,
 	\end{align*}
 	which shows that Condition (A3)  clearly holds.

\bibliographystyle{IEEEtran}
\bibliography{ref2}

\begin{thebibliography}{10}
\providecommand{\url}[1]{#1}
\csname url@samestyle\endcsname
\providecommand{\newblock}{\relax}
\providecommand{\bibinfo}[2]{#2}
\providecommand{\BIBentrySTDinterwordspacing}{\spaceskip=0pt\relax}
\providecommand{\BIBentryALTinterwordstretchfactor}{4}
\providecommand{\BIBentryALTinterwordspacing}{\spaceskip=\fontdimen2\font plus
\BIBentryALTinterwordstretchfactor\fontdimen3\font minus
  \fontdimen4\font\relax}
\providecommand{\BIBforeignlanguage}[2]{{%
\expandafter\ifx\csname l@#1\endcsname\relax
\typeout{** WARNING: IEEEtran.bst: No hyphenation pattern has been}%
\typeout{** loaded for the language `#1'. Using the pattern for}%
\typeout{** the default language instead.}%
\else
\language=\csname l@#1\endcsname
\fi
#2}}
\providecommand{\BIBdecl}{\relax}
\BIBdecl

\bibitem{pan2021}
J.~Pan, Y.~Li, and V.~Y.~F. Tan, ``Asymptotics of sequential composite
  hypothesis testing under probabilistic constraints,'' in \emph{IEEE
  International Symposium on Information Theory (ISIT)}, Melbourne, Australia,
  2021, pp. 172--177.

\bibitem{blahut1974hypothesis}
R.~Blahut, ``Hypothesis testing and information theory,'' \emph{IEEE
  Transactions on Information Theory}, vol.~20, no.~4, pp. 405--417, 1974.

\bibitem{tartakovsky2014sequential}
A.~Tartakovsky, I.~Nikiforov, and M.~Basseville, \emph{Sequential analysis:
  Hypothesis testing and changepoint detection}.\hskip 1em plus 0.5em minus
  0.4em\relax CRC Press, 2014.

\bibitem{neyman1933ix}
J.~Neyman and E.~S. Pearson, ``On the problem of the most efficient tests of
  statistical hypotheses,'' \emph{Philosophical Transactions of the Royal
  Society of London (Series A)}, vol. 231, pp. 289--337, 1933.

\bibitem{polyanskiy2014lecture}
Y.~Polyanskiy and Y.~Wu, ``Lecture notes on information theory,'' \emph{Lecture
  Notes for ECE563 (UIUC)}, 2014.

\bibitem{wald1948}
A.~Wald and J.~Wolfowitz, ``Optimum character of the sequential probability
  ratio test,'' \emph{Ann. Math. Statist.}, vol.~19, no.~3, pp. 326--339, 1948.

\bibitem{lalitha2016reliability}
A.~Lalitha and T.~Javidi, ``Reliability of sequential hypothesis testing can be
  achieved by an almost-fixed-length test,'' in \emph{IEEE International
  Symposium on Information Theory}.\hskip 1em plus 0.5em minus 0.4em\relax
  IEEE, 2016, pp. 1710--1714.

\bibitem{haghifam}
M.~Haghifam, V.~Y.~F. Tan, and A.~Khisti, ``Sequential classification with
  empirically observed statistics,'' \emph{IEEE Transactions on Information
  Theory}, vol.~67, no.~5, pp. 3095--3113, 2021.

\bibitem{zeitouni}
O.~Zeitouni, J.~Ziv, and N.~Merhav, ``When is the generalized likelihood ratio
  test optimal?'' \emph{IEEE Transactions on Information Theory}, vol.~38,
  no.~5, pp. 1597--1602, 1992.

\bibitem{lai2002asymptotic}
T.-L. Lai, ``Asymptotic optimality of generalized sequential likelihood ratio
  tests in some classical sequential testing problems,'' \emph{Sequential
  Analysis}, vol.~21, no.~4, pp. 219--247, 2002.

\bibitem{Li14}
Y.~{Li}, S.~Nitinawarat, and V.~V. Veeravalli, ``Universal outlier hypothesis
  testing,'' \emph{IEEE Transactions on Information Theory}, vol.~60, no.~7,
  pp. 4066--4082, 2014.

\bibitem{Li17}
Y.~Li, S.~Nitinawarat, and V.~V. Veeravalli, ``Universal sequential outlier
  hypothesis testing,'' \emph{Sequential Analysis}, vol.~36, no.~3, pp.
  309--344, 2017.

\bibitem{Xiaoou}
X.~Li, J.~Liu, and Z.~Ying, ``Generalized sequential probability ratio test for
  separate families of hypotheses,'' \emph{Sequential Analysis}, vol.~33,
  no.~4, pp. 539--563, 2014.

\bibitem{strassen1962asymptotische}
V.~Strassen, ``Asymptotische abschatzugen in {Shannon's} informationstheorie,''
  in \emph{Transactions of the Third Prague Conference on Information Theory
  etc, 1962. Czechoslovak Academy of Sciences, Prague}, 1962, pp. 689--723.

\bibitem{tan2014asymptotic}
V.~Y.~F. Tan, ``Asymptotic estimates in information theory with non-vanishing
  error probabilities,'' \emph{Foundations and Trends{\textregistered} in
  Communications and Information Theory}, vol.~11, no. 1-2, pp. 1--184, 2014.

\bibitem{yonglong}
Y.~{Li} and V.~Y.~F. {Tan}, ``Second-order asymptotics of sequential hypothesis
  testing,'' \emph{IEEE Transactions on Information Theory}, vol.~66, no.~11,
  pp. 7222--7230, 2020.

\bibitem{vaart_1998}
A.~W. {van der Vaart}, \emph{Asymptotic Statistics}.\hskip 1em plus 0.5em minus
  0.4em\relax Cambridge University Press, 1998.

\bibitem{wainwright2008graphical}
M.~J. Wainwright and M.~I. Jordan, \emph{Graphical models, exponential
  families, and variational inference}.\hskip 1em plus 0.5em minus 0.4em\relax
  Now Publishers Inc, 2008.

\bibitem{allan}
A.~R. Sampson, ``{Characterizing Exponential Family Distributions by Moment
  Generating Functions},'' \emph{The Annals of Statistics}, vol.~3, no.~3, pp.
  747 -- 753, 1975.

\bibitem{bierens_2004}
H.~J. Bierens, \emph{Modes of Convergence}, ser. Themes in Modern
  Econometrics.\hskip 1em plus 0.5em minus 0.4em\relax Cambridge University
  Press, 2004, pp. 137--178.

\bibitem{durrett2004probability}
R.~Durrett, \emph{Probability: Theory and Examples}.\hskip 1em plus 0.5em minus
  0.4em\relax Duxbury Press, 2004.

\bibitem{Cover}
T.~M. Cover and J.~A. Thomas, \emph{Elements of Information Theory (Wiley
  Series in Telecommunications and Signal Processing)}.\hskip 1em plus 0.5em
  minus 0.4em\relax USA: Wiley-Interscience, 2006.

\bibitem{amari}
S.-I. Amari and H.~Nagaoka, \emph{Methods of Information Geometry}, ser.
  Translations of Mathematical Monographs.\hskip 1em plus 0.5em minus
  0.4em\relax American Mathematical Society, 2007.

\bibitem{polyanskiy2010channel}
Y.~Polyanskiy, \emph{Channel Coding: Non-Asymptotic Fundamental Limits}.\hskip
  1em plus 0.5em minus 0.4em\relax Princeton University, 2010.

\bibitem{10.2307/1990053}
A.~C. Berry, ``The accuracy of the {G}aussian approximation to the sum of
  independent variates,'' \emph{Transactions of the American Mathematical
  Society}, vol.~49, no.~1, pp. 122--136, 1941.

\bibitem{boyd2004convex}
S.~Boyd and L.~Vandenberghe, \emph{Convex Optimization}.\hskip 1em plus 0.5em
  minus 0.4em\relax Cambridge University Press, 2004.

\bibitem{Spivak71}
M.~Spivak, \emph{Calculus On Manifolds: A Modern Approach To Classical Theorems
  Of Advanced Calculus}.\hskip 1em plus 0.5em minus 0.4em\relax Taylor \&
  Francis Inc, 1971.

\end{thebibliography}

\begin{IEEEbiographynophoto}{Jiachun Pan}
is currently a Ph.D.\ candidate in the Department of Electrical and Computer Cngineering in National University of Singapore (NUS). She received the B.S.\ degree from University of Electronic Science and Technology of China (UESTC) in 2015 and M.Eng.\ degree from University of Science and Technology of China (USTC) in 2019. Her research interests include information theory and statistical learning.
\end{IEEEbiographynophoto}

\vfill

\begin{IEEEbiographynophoto}{Yonglong Li}
is a research fellow at the Department of Electrical and Computer Engineering, National University of Singapore. He received the bachelor degree in Mathematics from Zhengzhou University in 2011 and the Ph.D.\ degree in Mathematics from the University of Hong Kong in 2015. From 2017 to 2019,  he was a postdoctoral fellow at the Center for Memory and Recording Research (CMRR), University of California, San Diego.
\end{IEEEbiographynophoto}
\vfill

\begin{IEEEbiographynophoto}{Vincent Y.\ F.\ Tan} (S'07-M'11-SM'15) was born in Singapore in 1981. He received the B.A.\ and M.Eng.\ degrees in electrical and information science from Cambridge University in 2005, and the Ph.D.\ degree in electrical engineering and computer science (EECS) from the Massachusetts Institute of Technology (MIT) in 2011. He is currently a Dean’s Chair Associate Professor with the Department of Electrical and Computer Engineering and the Department of Mathematics, National University of Singapore (NUS). His research interests include information theory, machine learning, and statistical signal processing.

Dr.\ Tan is a member of the IEEE Information Theory Society Board of Governors. He was an IEEE Information Theory Society Distinguished Lecturer from 2018 to 2019. He received the MIT EECS Jin-Au Kong Outstanding Doctoral Thesis Prize in 2011, the NUS Young Investigator Award in 2014, the Singapore National Research Foundation (NRF) Fellowship (Class of 2018), and the NUS Young Researcher Award in 2019. He is currently serving as a Senior Area   Editor for the IEEE \textsc{Transactions on Signal Processing} and for the IEEE \textsc{Transactions on Information Theory}.
\end{IEEEbiographynophoto}

\end{document}